\documentclass[preprint2]{aastex6}
\usepackage{amsmath} 
\usepackage{changepage}

\newcommand{\msun}{{\rm M_\odot}}
\newcommand{\rsun}{{\rm R_\odot}}
\newcommand{\lsun}{{\rm L_\odot}}
\newcommand{\zxsun}{{\rm (Z/X)_\odot}}
\newcommand{\tausun}{\tau_\odot}	
\newcommand{\yini}{Y_{\rm ini}}
\newcommand{\zini}{Z_{\rm ini}}
\newcommand{\ysur}{Y_{\rm S}}
\newcommand{\zsur}{Z_{\rm S}}
\newcommand{\ycen}{Y_{\rm C}}
\newcommand{\zcen}{Z_{\rm C}}
\newcommand{\rcz}{R_{\rm CZ}}
\newcommand{\alphamlt}{\alpha_{\rm MLT}}
\newcommand{\phib}{\Phi({\rm ^8B})}
\newcommand{\phibe}{\Phi({\rm ^7Be})}
\newcommand{\phin}{\Phi({\rm ^{13}N})}
\newcommand{\phio}{\Phi({\rm ^{15}O})}
\newcommand{\phif}{\Phi({\rm ^{17}F})}
\newcommand{\phipp}{\Phi({\rm pp})}
\newcommand{\phipep}{\Phi({\rm pep})}
\newcommand{\phihep}{\Phi({\rm hep})}

\definecolor{verdon}{cmyk}{1,0.5,1,0}

\begin{document}

\title{A new Generation of Standard Solar Models}

\author{N\'uria Vinyoles\altaffilmark{1},
 Aldo M. Serenelli\altaffilmark{1},
 Francesco L. Villante\altaffilmark{2,3},
 Sarbani Basu\altaffilmark{4},
 Johannes Bergstr\"om\altaffilmark{5},
 M.C. Gonzalez-Garcia\altaffilmark{5,6,7},
 Michele Maltoni\altaffilmark{8},
 Carlos Pe\~na-Garay\altaffilmark{9,10},
 Ningqiang Song\altaffilmark{7}}

\altaffiltext{1}{Institut de Ci\`encies de l'Espai (CSIC-IEEC), Campus UAB, Carrer de Can Magrans, S/N, E-08193 Barcelona, Spain}\email{Email to: vinyoles@ice.csic.es, aldos@ice.csic.es}
\altaffiltext{2}{Dipartimento di Scienze Fisiche e Chimiche, Universit\`a dell'Aquila, I-67100 L'Aquila, Italy}
\altaffiltext{3}{Istituto Nazionale di Fisica Nucleare (INFN), Laboratori Nazionali del Gran Sasso (LNGS), I-67100 Assergi  (AQ), Italy}
\altaffiltext{4}{Department of Astronomy, Yale University, PO Box 208101, New Haven, CT 06520, USA}
\altaffiltext{5}{Departament de F\'isica Qu\`antica i Astrof\'isica and ICC-UB,  Universitat de Barcelona, Av. Diagonal 647, E-08028 Barcelona, Spain.}
\altaffiltext{6}{Instituci\'o Catalana de Recerca i Estudis Avan\c{c}ats (ICREA), Pg. Llu\'is Companys 23, E-08010 Barcelona, Spain}
\altaffiltext{7}{C.N. Yang Institute for Theoretical Physics, SUNY at Stony Brook, Stony Brook,  NY 11794-3840, USA}
\altaffiltext{8}{Instituto de F\'isica T\'eorica UAM/CSIC, Calle de Nicol\'as Cabrera 13-15, Universidad Aut\'onoma de Madrid, Cantoblanco, E-28049 Madrid, Spain}
\altaffiltext{9}{Instituto de F\'isica Corpuscular, CSIC-UVEG, P.O. 22085, Valencia, E-46071, Spain}
\altaffiltext{10}{Laboratorio Subterr\'aneo de Canfranc, Estaci\'on de Canfranc, E-22880, Spain}

\begin{abstract}

We compute a new generation of standard solar models (SSMs) that includes recent updates on some important nuclear reaction rates and a more consistent treatment of the equation of state. Models also include a novel and flexible treatment of opacity uncertainties based on opacity kernels, required in the light of recent theoretical and experimental works on radiative opacity. Two large sets of SSMs, each based on a different canonical set of solar abundances with high and low metallicity (Z), are computed to determine model uncertainties and correlations among different observables. We present detailed comparisons of high- and low-Z models against different ensembles of solar observables including solar neutrinos, surface helium abundance, depth of convective envelope and sound speed profile. A global comparison, including all observables, yields a p-value of 2.7$\sigma$ for the high-Z model and 4.7$\sigma$ for the low-Z one. When the sound-speed differences in the narrow region of $0.65 < r/\rsun < 0.70$ are excluded from the analysis, results are 0.9$\sigma$ and 3.0$\sigma$ for high- and low-Z models respectively. These results show that: high-Z models agree well with solar data but have a systematic problem right below the bottom of the convective envelope linked to steepness of molecular weight and temperature gradients, and that low-Z models lead to a much more general disagreement with solar data. We also show that, while simple parametrizations of opacity uncertainties can strongly alleviate the solar abundance problem, they are insufficient to substantially improve the agreement of SSMs with helioseismic data beyond that obtained for high-Z models due to the intrinsic correlations of theoretical predictions. 
\end{abstract}

\keywords{}

\section{Introduction} \label{sec:intro}

Standard solar models (SSMs) have played a fundamental role in one of the most important discoveries in physics: neutrino flavor oscillations. After several decades during which the \emph{solar neutrino problem} puzzled both particle physicists and astrophysicists, helioseismology allowed to construct an accurate picture of the solar interior. The agreement between standard solar models and helioseismic results was astonishingly good, it provided both strong support to stellar (solar) evolution theory and to a particle physics solution for the problem of the missing neutrinos \citep{cd1996,bahcall98,bahcall01}. 

The solar surface composition, determined with spectroscopic techniques, is a fundamental constraint in the construction of SSMs. The development of three dimensional hydrodynamic models of the solar atmosphere, of techniques to study line formation under non-local thermodynamic conditions and the improvement in atomic properties (e.g. transition strengths) have led since 2001 to a complete revision of solar abundances \citep{ags05,agss09,caffau:2011}. There is no complete agreement among authors, and some controversy still remains as to what the best values for the new spectroscopic abundances are. However, there is consensus in that all determinations of the solar metallicity based on the new generation of spectroscopic studies yield a solar metallicity that is substantially lower than older spectroscopic results \citep{gn93,gs98}, in particular for the volatile and most abundant C, N, and O.

Almost all determinations of element abundances for astronomical objects rely upon solar abundances and the new solar abundances, particularly those from \citet{agss09}, have become a new standard. For modelling of the solar interior, however, they have brought about a series of problems because solar models in general, and SSMs in particular, based on the low-Z AGSS09 abundances fail to reproduce all helioseismic probes of solar properties. This \emph{solar abundance problem} \citep{basu04,bahcall05,delahaye:2006} has defied a complete solution. All proposed modifications to physical processes in SSMs offer, at best, only partial improvements in some helioseismic probes (e.g. \citealp{guzik:2005,castro07,basu08,guzik10,serenelli11}). An alternative possibility is to consider modifications to the physical inputs of SSMs at the level of the constitutive physics, radiative opacities in particular.

The effective opacity profile in the solar interior results from the combination of the reigning thermodynamic conditions, including composition, and the atomic opacity calculations at hand. Early works \citep{bahcall:2005,montalban:2004} already suggested that a localized increase in opacities could solve or, at least, alleviate the disagreement of low-Z solar models with helioseismology.  \citet{cd2009,villante10} have concluded that a tilted increase in radiative opacities, with a few percent increase in the solar core and a larger (15-20\%) increase at the base of the convective envelope could lead to low-Z SSMs that would satisfy helioseismic probes equally as well as SSMs based on the older, higher, metallicities.

Recent years have seen a surge of activity in theoretical calculations of atomic radiative opacities. Updated calculations \citep{Badnell05} by the Opacity Project have led the way, followed by OPAS \citep{blancard12,mondet:2015}, STAR \citep{krief:2016a} and a new version of OPLIB, the opacities from Los Alamos \citep{colgan:2016}. For conditions in solar interiors, all theoretical opacities agree with each other within 5\%. Interestingly, \citet{Bailey15} have presented the first ever measurement of opacity under conditions very close to those at the bottom of the solar convective envelope. While the experiment has been carried out only for iron, their conclusion is that all theoretical calculations predict a too low Rosseland mean opacity, at a level of $7\pm 4\%$, for the temperature and density combinations realized in the experiment. \citet{Krief16} have casted additional doubts in the accuracy of currently available opacity calculations after showing that the approximations done in the modelling of line broadening have a critical impact on the final Rosseland mean opacity. 

In parallel to work on radiative opacities, there have been new determinations of important nuclear reaction rates affecting energy and neutrino production in the Sun. These updates introduce differences in model expectations for neutrino fluxes  comparable to current experimental uncertainties of the well determined $\phibe$ and $\phib$ fluxes. Additionally, development  in \texttt{GARSTEC} (GARching STEllar Code; \citealp{GARSTEC}) allows now the implementation of an equation of state obtained for a mixture of elements that is consistent with the solar composition adopted in the calculation of the SSM. 

Motivated by the developments just described, in this work we present a new generation of SSMs that includes updates to the microscopic input physics of the models. Also, in the light of the new results in radiative opacities, we improve on our previous treatment of opacity uncertainties in solar model predictions. The implementation of a new flexible scheme based on opacity kernels \citep{Tripathy98} allows us now to test the impact of any opacity error function without the need to perform lengthy calculations of large sets of solar models. In addition to presenting the models, we carry out an exhaustive comparison of SSMs based on alternative compositions against different ensembles of solar observables. We further test the hypothesis that radiative opacities can in fact offer a solution to the solar abundance problem. 

The article is structured as follows. In \S~\ref{sec:SSMs} we detail the changes in the input of SSMs with respect to the previous generation of models and discuss the general approach to include input physics uncertainties in model calculations, in particular the implementation of opacity kernels to treat opacity uncertainties. In \S~\ref{sec:results} we present results of the new SSMs for helioseismic probes and solar neutrinos. It also contains  a complete discussion of how models of high- and low-Z compare to solar data, together with a discussion of the impact of correlations in models among different observables. This is an important aspect almost always missed in the literature, particularly when helioseismic probes are under consideration (see \citealt{villante14} for a counterexample). \S~\ref{sec:ssmerrors} gives a full discussion of uncertainties in SSMs: it summarizes our new Monte Carlo (MC) simulations, identifies dominant error sources using power-law expansions and considers the impact of different parametrizations of opacity uncertainties. In \S~\ref{sec:summary} we summarize our most relevant findings. 
Finally, all the detailed information such as the structure of the new solar models, distribution of production of solar neutrinos, results of MC simulations and update power-law dependences of neutrino fluxes on solar input parameters can be found online\footnote{\url{http://www.ice.cat/personal/aldos/Solar_Data.html}}.

\section{Standard Solar Models} \label{sec:SSMs}

SSMs are a snapshot in the evolution of a 1\,$\msun$ star, calibrated to match present-day surface properties of the Sun. In our models, two basic assumptions are that the Sun was initially chemically homogeneous and that at all moments during its evolution up to the present solar age $\tausun=4.57$\,Gyr mass loss is negligible. The calibration is done by adjusting the mixing length parameter ($\alphamlt$) and the initial helium and metal mass fractions ($\yini$ and $\zini$ respectively) in order to satisfy the constraints imposed by the present-day solar luminosity $\lsun$, radius $\rsun$, and surface metal to hydrogen abundance ratio $\zxsun$.  

In \cite{serenelli11} a generation of SSMs was computed using the nuclear reaction rates recommended in the Solar Fusion II paper (\citealp{sfII}; hereafter A11). For simplicity, we refer to them as the SFII SSMs. Here we present a new generation of SSMs, Barcelona 2016 or B16 for short. B16 models  share with the SFII models much of the physics. However, important changes include updates in some nuclear reaction rates and, most notably, a new treatment of uncertainties due to radiative opacities. The common features between generations of SSMs are described in the paragraph below. In Sect.\,\ref{sec:updates} we detail the changes in the input physics. There, we also discuss new results on important input physics for solar models that have nevertheless not made it into our final choice of SSM inputs, in particular solar composition and results on the ${\rm ^3He(^4He,\gamma)^7Be}$ rate. In Sect.\,\ref{sec:uncertainties} we discuss the treatment of errors, with special emphasis in radiative opacities. 

Models have been calculated using \texttt{GARSTEC}. Physics included in the models, unless stated otherwise, is as follows. Atomic opacities are from OP \cite{Badnell05} and are complemented at low temperatures with molecular opacities from \citet{ferguson05}. Microscopic diffusion of helium and metals is followed according to the method of \citet{thoul94}. Convection is treated according to the mixing length theory and the atmosphere is a grey Krishna-Swamy model \citep{ks66}. For more details about SSMs calibrations and physics included in the models see \citet{serenelli16}.

\subsection{B16 - a new generation of SSMs}
\label{sec:updates}

Here we describe in some detail the differences between the older SFII and the new B16 generations of SSMs and discuss the reasons why some updates in important SSM inputs have not been included in our new SSM calculations.

\textbf{Composition: } Solar photospheric (surface) abundances of almost all metals can be determined from spectroscopy. In the context of solar models, the only relevant exceptions are Ne and Ar, the latter with a much lesser influence, that have to be determined by other, more indirect, methods (see \citealt{agss09} for details). For refractories, however, meteorites offer a very valuable alternative method  (see e.g. \citealt{lodders:09}) and, in fact, elemental abundances determined from meteorites have been historically more robust than spectroscopic ones. 

In \citet{agss09} mild, but worth of attention, differences existed between photospheric and meteoritic abundances for some refractory elements. In particular, photospheric Fe and Ca were 0.05\,dex higher than their meteoritic counterparts, while for  Mg the difference was 0.07\,dex. In a recent series of papers \citet{scott2,scott1,scott3} (for short, AGSS15) have updated previous results from AGSS09 for all but the CNO elements (and Ne, as its abundance is linked to that of O). Interestingly, the photospheric abundances of the discrepant elements just mentioned have shifted towards meteoritic values and with the new AGSS15 recommended photospheric values the differences above are now 0.02\,dex for Fe, 0.03\,dex for Ca and 0.06\,dex for Mg. 

In the past, the robustness of meteoritic abundances has made them our preferred choice as source of solar abundances. Recent changes in the AGSS15 photospheric values compared to AGSS09 give added strength to this preference. Then, in building our SSMs, the sets of solar abundances we use are always composed by meteoritic values for refractory elements and photospheric values for volatile elements. Both scales are tied together by forcing a rigid translation of the meteoritic scale such that the meteoritic abundance of Si matches the photospheric value.

The photospheric abundance of Si in AGSS15 remained unchanged from AGSS09. For this reason, the discussion above on our choice of a combined photospheric and meteoritic solar mixture, and the fact that unfortunately AGSS15 does not include a revision of CNO abundances,  does not lead to any changes with respect to our AGSS09 based set of solar abundances. Then the two central sets of solar abundances we use in this work are the same employed in \citet{serenelli11}:

\begin{itemize}

\item GS98: Photospheric (volatiles) + meteoritic (refractories) abundances from \citet{gs98}. The metal-to-hydrogen ratio used for the calibration is  $\zxsun = 0.0229$,


\item AGSS09met: Photospheric (volatiles) + meteoritic (refractories) abundances from \citet{agss09}. The metal-to-hydrogen ratio used for the calibration is  $\zxsun = 0.0178$.

\end{itemize}

Individual abundances of the most relevant elements for solar modeling for the different compositions we use are given in Table\,\ref{tab:compo}.

\begin{table}
\centering
\begin{tabular}{c| c c}
 Element &  GS98 & AGSS09met \\
\hline
 C & $8.52  \pm 0.06$ & $8.43 \pm 0.05$ \\
 N & $7.92  \pm 0.06$ & $7.83 \pm 0.05$ \\
 O & $8.83  \pm 0.06$ & $8.69 \pm 0.05$ \\ 
 Ne & $8.08 \pm 0.06$ & $7.93 \pm 0.10$ \\
 Mg & $7.58 \pm 0.01$ & $7.53 \pm 0.01$ \\
 Si & $7.56 \pm 0.01$ & $7.51 \pm 0.01$ \\
 S  & $7.20 \pm 0.06$ & $7.15 \pm 0.02$ \\
 Ar & $6.40 \pm 0.06$ & $6.40 \pm 0.13$ \\
 Fe & $7.50 \pm 0.01$ & $7.45 \pm 0.01$ \\
\hline
$\zxsun$ & 0.02292 & 0.01780 \\ \hline 
\end{tabular}
\caption{Abundances of the \citet{gs98} and \citet{agss09} solar mixtures used in this work  given as $\log{\epsilon_i} \equiv \log{N_i/N_H} + 12$. Only elements that most strongly contribute to uncertainties in SSM modeling are included.
\label{tab:compo}}
\end{table}

\textbf{Equation of state:} SFII models used the equation of state (EoS) by OPAL \cite{opaleos:2001} in its 2005 version. This EoS has one slight disadvantage: the mixture of metals includes only C, N, O and Ne and their relative abundances are hardwired in the tables provided and cannot be modified. This does not represent an obstacle in using the OPAL EoS which has been, in fact, the most widely used EoS for solar models. However, it is desirable that the EoS offers consistency with the metal mixture used in the calibration of the solar model.

FreeEOS, the EoS developed by A. Irwin \citep{cassisi03}, allows us to overcome this difficulty. The source codes are freely available\footnote{\url{http://freeeos.sourceforge.net/}} and although the running time is too long to allow \emph{inline} implementation of the EoS during solar model calculations, precomputed tables with any desired solar composition can be computed in advanced. 
FreeEOS and OPAL EoS are thermodynamically consistent and in good agreement for the solar case. We have performed a comparison between the two EoS, with the composition fixed to the six element mixture hardwired in OPAL EoS, and found differences smaller than  $0.1\%$ for the pressure or $0.02\%$ for $\Gamma_1$. Differences for the sound speed are about $0.05\%$ at the center and they decrease to about $0.02\%$ in the intermediate region; these differences are well below the sound speed errors considered in this work (see Fig. \ref{fig:dcerror}). Extensive comparisons between FreeEOS and OPAL EoS are available at \url{http://freeeos.sourceforge.net/documentation.html}.

Due to its flexibility, and its excellent performance when compared to OPAL EoS, we adopt FreeEOS as our standard EoS for the B16 SSMs. For the first time, EoS tables calculated consistently for each of the compositions used (GS98 and AGSS09met) are used in the solar calibrations. This is a qualitative step forward albeit quantitatively differences in the predicted solar properties by use of consistent EoS tables are small and have minimal impact in the production of solar neutrinos or helioseismic diagnostics used in this work. This can be, nevertheless, much more important in the context of abundance determinations from EoS features such as the depression of the adiabatic index $\Gamma_1$ (see e.g. \citealp{lin:2007,vorontsov:2013}). In fact, we find that two SSMs calibrated using the same solar mixture but using EoS tables computed with GS98 or AGSS09met mixtures respectively, lead to differences in $\Gamma_1$ of the order of $2\times 10^{-4}$ in the region $r/\rsun > 0.7$. This is comparable to results shown in \citet{lin:2007} using a different EoS, and it is about two times larger than the error in the helioseismic determination of $\Gamma_1$. Further study of the impact of the EoS on $\Gamma_1$ are beyond the scope of this paper.

\textbf{Nuclear rates:} The most relevant changes in the B16 SSMs compared to SFII models arise from updates in the nuclear reaction rates. As usual, the astrophysical  $S$-factor $S(E)$ is expressed as a Taylor series around $E=0$ (Eq.\,\ref{eq:sfactor}, see also \citealt{sfII})
\begin{equation}
S(E)=S(0) + S'(0)\cdot E + \frac{1}{2} S''(0) \cdot E^2 + \mathcal{O}(E^3).
\label{eq:sfactor}
\end{equation}
Updates in the reaction rates are generally introduced as changes in $S(0)$ and, eventually, the first and higher order derivatives. New $S(0)$ values and errors are summarized in Table~\ref{tab:params1} together with the fractional changes in $S(0)$ with respect to A11.  Rates not listed in Table\,\ref{tab:params1} are taken from A11 and remain unchanged with respect to the SFII SSMs.

\begin{table}
\centering
\begin{tabular}{c| r c  c c}
 & \multicolumn{1}{c}{$S(0)$} & Uncert.\,\%  & $\Delta S(0)/S(0)$ &  Ref.\\
\hline
$\rm{S_{11}}$ & $4.03\cdot 10^{-25}$ & 1  & 0.5\%$^\dagger$ & 1,2,3\\
$\rm{S_{17}}$ & $2.13\cdot 10^{-5}$  & 4.7 & +2.4\% & 4\\
$\rm{S_{114}} $ & $1.59\cdot 10^{-3}$ & 7.5 & -4.2\% & 5\\ \hline
\end{tabular}
\caption{Astrophysical S-factors (in units of MeV\,b) and uncertainties updated in this work. Fractional changes with respect to A11 are also included. $^\dagger S_{11}(0)$ underestimates the actual increase in $S_{11}(E)$ that is dominated by changes in higher orders in the Taylor expansion (see text). (1)~\citet{marcucci13}, (2)~\citet{tognelli15}, (3) \citet{acharya:2016}, (4)~\citet{nollett15}, (5)~\citet{Marta11}. \label{tab:params1}}
\end{table}

Details about the updated rates are specified below:

\begin{itemize}
 
\item $\boldsymbol{{\rm p(p,e^+\nu_e)d}}$: $S_{11}(E)$ has been recalculated in \citet{marcucci13} by using  chiral effective field theory framework, including the P-wave contribution that had been previously neglected. In addition, they provide fits to $S(E)$ using the Taylor expansion (Eq.\,\ref{eq:sfactor}). For the leading order they obtain $S_{11}(0)= (4.03 \pm 0.006) \cdot 10^{-25}\,\rm{MeV\,b}$. This is 0.5\% higher and with a much smaller error than the recommended value in A11. More recently, and also using chiral effective field theory, \citet{acharya:2016} determined $S_{11}(E)$, resulting in $S_{11}(0)= 4.047^{+0.024}_{-0.032}\cdot 10^{-25}\,\rm{MeV\,b}$. This is in very good agreement with \citet{marcucci13} result. \citet{acharya:2016} have performed a more thorough assessment of uncertainty source leading to an estimated error of 0.7\%, much closer to the 1\% uncertainty which was obtained by A11. Based on the larger error estimate by \citet{acharya:2016} and the difference in the central values for $S_{11}(0)$ and higher order derivatives between both works, we prefer to make a conservative choice and adopt a 1\% uncertainty for the p+p reaction rate.

The evaluation of $S(E)$ presented by \citet{marcucci13} allows a full integration of the p+p rate, avoiding the Taylor expansion. Moreover, \citet{tognelli15} provide a routine\footnote{\url{http://astro.df.unipi.it/stellar-models/pprate/}} to directly compute the p+p rate that we use to implement this rate in \texttt{GARSTEC}. In Fig.\,\ref{fig:ppratio} we plot as a function of temperature the ratio between the newly adopted \citep{marcucci13} and the older A11 reaction rates. For comparison, the rate inferred from \citet{acharya:2016} is also shown. The comparison with the A11 rate shows a larger variation than the 0.5\% difference quoted for $S_{11}(0)$ that is due to changes in the first and the higher order derivatives, as well as to the different integration methods to compute the rate. The solid black line in the plot shows the distribution of p+p reactions in the Sun and illustrates that for solar conditions the average difference in the rates is of order 1.3\%.

\begin{figure}
	\centering 
	\includegraphics[width=0.95\columnwidth]{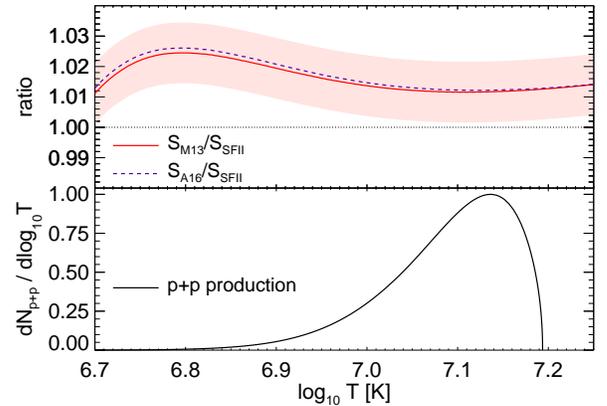}
        \caption{Top panel: ratio of p+p reaction rates as a function of temperature between \citet{marcucci13} and A11 (red solid line) and between \citet{acharya:2016} and A11 (purple dashed line). The band is our conservative 1$\sigma$ uncertainty (1\%). 
        Bottom panel: number of p+p reactions produced per $\delta \log{T}$ interval as a function of temperature in SSMs (arbitrary units). \label{fig:ppratio}}
\end{figure}

\item $\boldsymbol{{\rm ^7Be(p,\gamma)^8B}}$: A11 recommends $S_{17}(0) = (2.08 \pm 0.07 \pm 0.14 )\cdot 10^{-5} \,\rm{MeV\,b}$, where the first error term comes from uncertainties in the different experimental results and the second one from considering different theoretical models employed for the low-energy extrapolation of the rate. \citet{nollett15} present a new low-energy extrapolation  based on Halo Effective Field Theory, which allows for a continuous parametric evaluation of all low-energy models. Marginalization over the family of continuous parameters then amounts to marginalizing the results over the different low-energy models. They obtain $S_{17}(0) = (2.13 \pm 0.07)\cdot 10^{-5} \,\rm{MeV\,b}$. We note  the uncertainty equals that of the experimental uncertainty as given above. In A11, this error source was inflated to accommodate systematic differences seen among different experimental results (see in particular their Appendix on Treating Uncertainties). This problem was not found by \citet{nollett15} in their analysis. While the different findings by A11 and \citet{nollett15} regarding inconsistency in the nuclear data need further study (K.M. Nollet 2016, private communication), we prefer to err on the safe side and adopt an intermediate error between those from \citet{nollett15} and A11. Therefore, we finally adopt $S_{17}(0) = (2.13 \pm 0.1)\cdot 10^{-5} \,\rm{MeV\,b}$ and add the two error sources quadratically. We also have updated the derivatives by using the recommended values of \cite{nollett15}.\\

\item $\boldsymbol{{\rm ^{14}N(p,\gamma)^{15}O}}$: \citet{Marta11} present new cross-section data for this reaction obtained at the Laboratory for Underground Nuclear Astrophysics (LUNA) experiment. With the new data and using R-matrix analysis they recommend a new value for the ground-state capture of $S_{GS}(0) = (0.20 \pm 0.05 ) \cdot 10^{-3} \hspace{1mm}\rm{MeV\,b}$, down from the previously recommended value of $0.27 \cdot 10^{-3} \hspace{1mm}\rm{MeV\,b}$ (A11). Combined with other transitions (see Table XI in that work) this leads to  $S_{114}(0) = (1.59 \cdot 10^{-3}) \hspace{1mm}\rm{MeV\,b}$, about 4\% lower than the previous A11 recommended value. The derivatives and the errors remain unchanged.\\

\item $\boldsymbol{{\rm ^3He(^4He,\gamma)^7Be}}$: \citet{deboer14} combine \emph{recent} (post 2004) experimental results including those in A11 but also newer data at medium and higher energies, from 300 up to 3500 keV (see references in their work). They perform a global \emph{R}-matrix fit to determine the extrapolated $S_{34}(0)$ value and a Monte Carlo analysis of  the \emph{R}-matrix fit to determine the uncertainties in the rate.  This is a very different approach to that used in A11, where microscopic models were used to determine $S_{34}(0)$ for four independent datasets and then combined statistically to provide a final result for $S_{34}(0)$. The final value reported by \citet{deboer14} is  $S_{34}(0) = (5.42 \pm 0.23)\cdot 10^{-4} \,\rm{MeV\,b}$. The central value is  $\sim 3\%$ lower than the previous A11 recommended value. It should be pointed out that this value is systematically lower than three out of the four results used in A11 and very similar to the fourth one. The underlying reasons for these differences are not discussed in \citet{deboer14}. 

More recently, \citet{iliadis:2016} have performed a global Bayesian estimation of $S_{34}$ using the same data as in A11, but extended up to 1.6 MeV instead of 1 MeV, and found $S_{34}(0)= (5.72 \pm 0.12)\cdot 10^{-4} \,\rm{MeV\,b}$. This is 2\% higher than A11 and almost 6\% larger than \citet{deboer14}. The use of larger energy ranges in \citet{deboer14} and \citet{iliadis:2016} compared to what is required for an accurate determination of $S_{34}(E)$ in the energy range required for solar neutrino calculations (see A11) makes us wary of the impact that derivatives of $S(E)$ could have on the expansion. 

Given the reasons above, and that \citet{deboer14} and \citet{iliadis:2016} results bracket that from A11, we have decided to keep the latter as our preferred choice in B16 SSMs. We point out, however, that a reduction in $S_{34}(0)$ such as that claimed by \citet{deboer14} would have an impact in the comparison between solar neutrino data and SSMs built with the GS98 or AGSS09met compositions for the $\phib$ and $\phibe$ fluxes. We briefly touch upon this in Sect.\,\ref{sec:neutrinos}.

\end{itemize}

We use Salpeter's formulation of weak screening \citet{salpeter:1954}. The validity of this formulation for solar conditions, where electrons are only weakly degenerate, has been discussed in detail in \citet{gruzinov98}, where a more sophisticated approach was shown to lead, to within differences of about 1\%, to Salpeter's result. Other proposed deviations from this formulation have been discussed at length in \citet{bahcall02}, including different approaches to dynamic screening, and shown to be flawed or not well physically motivated. We therefore assume that uncertainties associated to electron screening are negligible in comparison to others entering SSM calculations. We are aware, however, that more recent calculations of dynamic screening \citep{mao:2009,mussack:2011} still leave some room for discussion on this topic.

To conclude this section, we present in Table\,\ref{tab:params2} a summary of other main cross sections and input parameters used to construct the SSMs.

\begin{table}[h]
\centering
\begin{tabular}{c| c c c}
Qnt. & Central value & $\sigma$ (\%) & Ref. \\
\hline
${\rm ^3He(^3He,2p)^4He}$ & 5.21 MeV\,b & 5.2 & 1 \\
${\rm ^3He(^4He,\gamma)^7Be}$ & $5.6\cdot 10^{-4}$ MeV\,b & 5.2 & 1\\
${\rm ^7Be(e^-,\nu_e)^7Li}$ & Eq (40) SFII & 2.0  & 1 \\
${\rm ^3He(p,e^+\nu_e)^4He}$ & $8.6 \cdot 10^{-20}$ MeV\,b & 30.2 & 1 \\
${\rm ^{16}O(p,\gamma)^{17}F} $ & $1.06 \cdot 10^{-2}$ MeV\,b & 7.6  & 1 \\
$\tausun$ & $4.57 \cdot 10^9$ yr& 0.44  & 2\\
diffusion & $1.0$ & 15.0&  2\\
$\lsun$ & $3.8418 \cdot 10^{33}$ $\rm{erg\,s^{-1}}$ & 0.4 & 2 \\ \hline
\end{tabular}
\caption{Central values for the main input parameters and the correspondent standard deviation. (1) A11, (2) \citet{bahcall06}.
\label{tab:params2}}
\end{table}

\subsection{Treatment of model uncertainties} \label{sec:uncertainties}

Within the framework defined by SSMs, the treatment of model uncertainties is generally simple. Most of the input physics in the models can be characterized by simple numbers such as the astrophysical factors or the surface abundance of a given element, as discussed in previous section. Tables~\ref{tab:compo}-\ref{tab:params2} list the uncertainties we adopt for each of the input quantities that allow such simple parametrizations. The adopted uncertainty for the microscopic diffusion coefficients deserves a special comment, as the coefficients cannot be obtained experimentally. The quoted 15\% uncertainty, the same used in previous SSM calculations by our group, stems from results presented in \citet{thoul94}, where the complete solution of the multiflow Burgers equations was initially presented in the context of SSMs. As discussed in that work, the uncertainty in the diffusion coefficients comes from the calculation of the Coulomb collision integrals. The comparison of their results with equivalent calculations available in the literature \citep{michaud:1991} yielded a $< 15\%$ difference in the diffusion coefficients for all relevant elements in the solar interior. The adopted 15\% uncertainty (1$\sigma$) is therefore conservative in more than one aspect: it is based on the difference between calculations (we could equally well define 1$\sigma$ as half the difference between calculations), and it reflects the largest difference between different calculations for all the solar interior and all relevant chemical elements. Later works showed that inclusion of radiative levitation, for instance \citep{turcotte:1998}, or quantum corrections to the collision integrals \citet{schlattl:2003}, have very minor effects in solar model calculations which are well within the adopted uncertainties.

A fundamentally important physical ingredient in solar models that cannot be quantified by just one parameter is the radiative opacity, which is a complicated function of temperature ($T$), density ($\rho$) and chemical composition ($X_i$) of the solar plasma. The magnitude and functional form of its uncertainty is currently not well constrained in available opacity calculations. As a result, representation of the uncertainty in radiative opacity by a single parameter \citep{serenelli13} or by taking the difference between two alternative sets of opacity calculations \citep{bahcall06,villante14} are  strong simplifications, at best. In this paper, instead, we choose to follow a general and flexible approach based on opacity kernels originally developed by \citet{Tripathy98} and later on by \citet{villante10}, which we describe next.

The reference opacities $\bar\kappa(\rho,T,X_i)$ in SSMs can be modified by a generic function of T, $\rho$ and $X_i$. For simplicity, we assume that opacity variations are parametrized as a function of $T$ alone such that
\begin{equation}
\kappa(\rho,\,T,\,X_{\rm   i}) = 
\left[1+\delta \kappa(T) \right] \, \overline{\kappa}(\rho,\,T,\,X_{\rm  i})
\label{prescription} 
\end{equation}
where $\delta \kappa(T)$ is an arbitrary function. The Sun responds linearly even to relatively large opacity variations $\delta \kappa (T)$ as shown by \citet{Tripathy98} and \citet{villante10}. 
Thus, the fractional variation of a generic SSM prediction
\begin{equation}
\delta Q \equiv Q/\bar Q - 1
\end{equation}
where $Q$ ($\bar Q$) corresponds to the modified (reference) value, can be described as
\begin{equation}
 \delta Q = \int \frac{dT}{T}K_Q (T) \delta \kappa (T)
 \label{eq:dqkappa}
\end{equation}
by introducing a suitable kernel $K_Q(T)$ that describes the response of the considered quantity to changes in the opacity at a given temperature. We determine the kernels $K_Q(T)$ numerically by studying the response of solar models to localized opacity changes as it was done in \citet{Tripathy98}. Our results agree very well in all cases except for variations in the chemical composition because our models include gravitational settling. However, this has a negligible effect for the calculation of opacity uncertainties. 

The evaluation of $\delta Q$ is subject to the choice we make for $\delta \kappa(T)$. In \citet{haxton:2008} and \citet{serenelli13} the opacity error was modelled as a 2.5\% constant factor at $1\sigma$ level, comparable to the maximum difference between the OP and OPAL \citep{opal} opacities in the solar radiative region. \citet{villante10} showed that this prescription underestimates the contribution of opacity uncertainty to the sound speed and convective radius error budgets because the opacity kernels for these quantities are not positive definite and integrate to zero for $\delta \kappa(T) = {\rm const}$. Later on, \citet{villante14} considered the temperature-dependent difference between OP and OPAL opacities as $1\sigma$ opacity uncertainty. However, it is by no means clear that this difference is a sensible measure of the actual level of uncertainty in current opacity calculations. 

Based on the previous reasons, here we follow a different approach inspired by the most recent experimental and theoretical results and some simple assumptions. The contribution of metals to the radiative opacity is larger at the bottom of the convective envelope ($\sim 70\%$) than at the solar core ($\sim 30\%$).  Also, \citet{Krief16} in a recent theoretical analysis of line broadening modelling in opacity calculations have found that uncertainties linked to this are larger at the base of the convective envelope than in the core. These arguments suggest that opacity calculations are more accurate at the solar core than in the region around  the base of the convective envelope. It is thus natural to consider error parameterizations that allow opacity to fluctuate by a larger amount in the external radiative region than in the center of the Sun. 

Taking all this into account, we consider the following parameterization for the opacity change $\delta \kappa(T)$:
\begin{equation}
\delta \kappa(T) = a + b \; \frac{\log_{10}(T_{\rm C}/T)} {\Delta}
\label{eq:dkappaerror}
\end{equation}
where $\Delta = \log_{10}(T_{\rm C}/T_{\rm CZ}) = 0.9$. $T_{\rm C}$ and $T_{\rm CZ}$ are the temperatures at the solar center and at the bottom of the convective zone respectively. This equation is applied only up to the lower regions of the convective envelope, where convection is adiabatic and changes in the opacity are irrelevant. Changes in the opacity in the uppermost part of the convective envelope and atmosphere are absorbed in the solar calibration by changes in the mixing length parameter and in sound speed inversions by the surface term and, in the context of SSMs, will not produce changes in the solar properties considered in the present work. By changing the parameters $a$ and $b$, one is able to rescale and tilt the solar opacity profile by arbitrary amounts. We consider them as independent random variables with mean equal to zero and dispersions $\sigma_a$ and $\sigma_b$, respectively. This corresponds to assuming that the opacity error at the solar center is $\sigma_{\rm in} = \sigma_a$, while it is given by $\sigma_{\rm out} \simeq \sqrt{\sigma^2_a + \sigma^2_b}$ at the base of the convective zone. We fix $\sigma_{\rm in} =\sigma_a = 2\%$  which is the average difference of the OP and OPAL opacity tables. This is also comparable to differences found with respect to the new Los Alamos OPLIB opacity tables \citep{colgan:2016}. For $\sigma_{\rm out}$ we choose 7\% (i.e. $\sigma_b=6.7\%$), motivated by the recent experimental results of \citet{Bailey15} that have measured the iron opacity at conditions similar to those at the base of the solar convective envelope and have found a $7\% \pm 4 \%$ increase with respect to the theoretical expectations. We note that the adopted functional form, see Eq.\,\ref{eq:dkappaerror}, for the opacity error function is a simplified parametric description of a more complex (and unknown) behaviour.
Our choice is motivated by practical reasons and by the important fact that \citet{cd2009} and \citet{villante10} have shown that an opacity solution to the solar abundance problem requires a tilt of opacity profile of the Sun by increasing opacity by a few percent at the solar center and a much larger increase (up to 15 to 20\%) at the base of the convective region, i.e. the kind of behaviour for $\delta \kappa(T)$ described by Eq.\,\ref{eq:dkappaerror} when $b\neq 0$.
The adopted functional form can mimic, moreover, the uncertainty in theoretical calculations introduced by line broadening modelling discussed by \citet{Krief16}.

\section{Results}
\label{sec:results}

Here we present the main results of the B16 SSMs for  GS98 and AGSS09met compositions and discuss differences with respect to our previous SFII models. Table~\ref{tab:ssmres} presents a summary of the most relevant quantities linked to the calibration of SSMs and helioseismic quantities. In Table~\ref{tab:chi2}   we quantify the agreement between SSMs and solar data and Table~\ref{tab:neutrinos} gives results for solar neutrino fluxes. Model errors and theoretical correlations among observable quantities have been obtained from MC simulations that are discussed to some detail in Section\,\ref{sec:ssmerrors}. 

\subsection{Helioseismology}\label{sec:helio}

Two helioseismic quantities widely used in assessing the quality of SSMs are the surface helium abundance $\ysur$ and the location of the bottom of the convective envelope $\rcz$. Both are listed in Table~\ref{tab:ssmres} together with the corresponding seismic  values. The model errors associated to these quantities are larger in B16 models than previously computed \citet{bahcall06} generations of SSMs because of the different treatment of uncertainties in radiative opacities (see Section\,\ref{sec:ssmerrors} for details). Compared to SFII models, we find a small decrease in the predicted $\ysur$ by 0.0003 for both compositions and a decrease in the theoretical $\rcz$ by 0.0007~$\rsun$, also for both compositions. The Pearson correlation between these quantities in SSMs is $\rho(\ysur,\rcz)=-0.35$  and $-0.41$ for B16-GS98 and B16-AGSS09met models respectively, as obtained from the MC calculations. A comparison of models and data for these two quantities yields $\chi^2=0.91$ and $\chi^2=6.45$ for GS98 and AGSS09met compositions that translate into 0.5$\sigma$ and 2.1$\sigma$ differences between models and data. This is summarised in Table~\ref{tab:chi2}.

\begin{table}[h]
\centering
\setlength{\tabcolsep}{2.5pt}
\begin{tabular}{l|ccc}
Qnt. & B16-GS98 & B16-AGSS09met & Solar \\
\hline
$\ysur$& $0.2426 \pm 0.0059 $&$0.2317 \pm 0.0059 $& $0.2485 \pm 0.0035$ \\
$\rcz/\rsun$& $0.7116 \pm 0.0048$ & $0.7223  \pm 0.0053$ & $0.713 \pm 0.001 $\\
$\langle\delta c/c \rangle$& $0.0005 ^{+0.0006}_{-0.0002}$ &$0.0021 \pm 0.001$ & 0$^a$ \\ \hline
$\alphamlt$ & $2.18 \pm 0.05$ &$2.11 \pm 0.05 $& - \\
$\yini$& $0.2718 \pm 0.0056 $ &$0.2613 \pm 0.0055 $& - \\
$\zini$ & $0.0187 \pm 0.0013$ &$0.0149 \pm 0.0009 $& - \\
$\zsur$& $0.0170 \pm 0.0012 $&$0.0134 \pm 0.0008 $& - \\
$\ycen$& $0.6328 \pm 0.0053 $ &$0.6217 \pm 0.0062 $& -\\
$\zcen$& $0.0200 \pm 0.0014 $ &$0.0159 \pm  0.0010 $& - \\ \hline
\end{tabular}
\caption{Main characteristics for the different SSMs with the correspondent model errors and the values for the observational values (when available) and their error. The observational values of $\ysur$ is taken from \citet{basu04} and $\rcz$ from \citet{basu97}. $^a$The solar value is zero, by construction of $\delta c/c$.}
\label{tab:ssmres}
\end{table}

\begin{figure}
	\centering 
	\includegraphics[width=0.95\columnwidth]{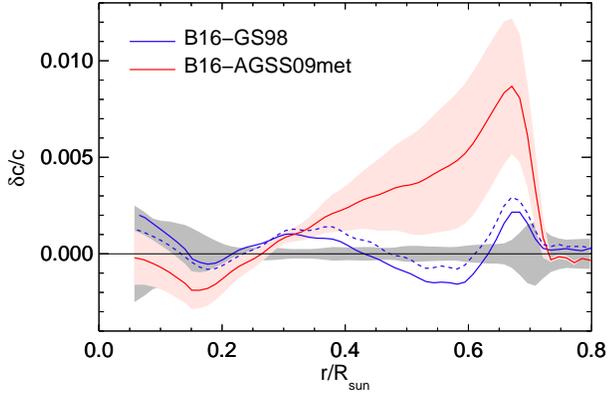}
        \caption{Fractional sound speed difference in the sense $\delta {\rm c /c = (c_\odot - c_{\rm mod})/c_{\rm mod}}$. Grey shaded regions corresponds to errors from the inversion procedure (see text for details). Red shaded region corresponds to errors from the model variation which we chose to plot around the AGSS09met central value (solid red line). An equivalent relative error band holds around the central value of the GS98 central value (solid blue line) which we do not plot for the sake of clarity. Dashed line shows, for comparison, results for the older SFII-GS98 SSM. \label{fig:ssmcs}}
\end{figure}

Fig.\,\ref{fig:ssmcs} shows the fractional sound speed difference as a function of solar radius. The solar sound speed differences have been obtained for each of the two SSMs by performing new sound speed inversions, using the appropriate reference solar model, based on the  BiSON-13 dataset (a combination of BiSON+MDI data) as described in \citet{basu:2009}. The resulting $\delta c/c$ curves are not too different with respect to SFII models.  This is expected because the differences between the two generations of models are mostly due to changes in the nuclear reaction rates. All rates have a negligible impact on the solar sound speed profile except for the ${\rm p(p,e^+\nu_e)d}$ rate. It is the newly adopted rate for this reaction that introduces the small differences with respect to the older generation of SSMs. This is shown in Fig.\,\ref{fig:ssmcs} by including the sound speed difference for the previous SFII-GS98 SSM in dashed line. A change in the p+p rate leads to structural changes in the structure of the model that are non-local \citep{villante10} due to the constraints imposed in building a SSM, in particular $\rsun$ and $\lsun$,  so the model sound speed is also affected at larger radii, where nuclear burning is negligible.

A quantitative assessment of the agreement between model and solar sound speeds is not straightforward. It requires a proper evaluation of model errors and correlations. Also, given a set of observed frequencies, extraction of the sound speed profile is sensitive to uncertainties in the measured frequencies, numerical parameters inherent to the inversion procedure and the solar model used as a reference model for performing the inversion. Such detailed analysis was carried out in \citet{villante14}, in which the SSM response to varying input parameters was modelled using power-law expansions and the three uncertainties related to the extraction of $\delta c/c$ from observed data were taken directly from \citet{deglinnoccenti:1997}.

In this work, we use large MC sets of SSMs  (Sect.\,\ref{sec:ssmerrors}) to account for model errors and correlations instead of using power-law expansions around a reference model. The total error from all input parameters in SSMs is illustrated in Fig.\,\ref{fig:ssmcs} as the shaded area embracing the B16-AGSS09met curve. Note that in comparison to previous estimates, e.g. \citet{villante14}, errors are larger due to the adoption of the larger opacity uncertainty. It should also be noted that model errors are strongly correlated across the solar radius. 

\begin{table}[h]
\centering
\setlength{\tabcolsep}{2.5pt}
\begin{tabular}{cccccc}
\multicolumn{2}{c}{} &   \multicolumn{2}{c}{GS98} & \multicolumn{2}{c}{AGSS09met}\\ \hline
Case & \multicolumn{1}{|c|}{dof}  &  $\chi^2$ & p-value\,$(\sigma)$ & $\chi^2$ & p-value\,$(\sigma)$ \\ \hline
$\ysur + \rcz$ only & \multicolumn{1}{|c|}{2} & 0.9 & 0.5 & 6.5 & 2.1 \\
$\delta c/c$ only  & \multicolumn{1}{|c|}{30} & 58.0 & 3.2 & 76.1 & 4.5 \\ 
$\delta c/c$ no-peak  & \multicolumn{1}{|c|}{28} & 34.7 & 1.4 & 50.0 & 2.7 \\ 
$\phibe+\phib$  & \multicolumn{1}{|c|}{2} & 0.2 & 0.3 & 1.5 & 0.6 \\
all $\nu$-fluxes  & \multicolumn{1}{|c|}{8} & 6.0 & 0.5 & 7.0 & 0.6 \\
\hline
global & \multicolumn{1}{|c|}{40} & 65.0 & 2.7 & 94.2 & 4.7 \\
global no-peak & \multicolumn{1}{|c|}{38} & 40.5 & 0.9 & 67.2 & 3.0 \\
\hline
\end{tabular}
\caption{Comparison of B16 SSMs against different ensembles of solar observables. \label{tab:chi2}}
\end{table}

\begin{figure}
	\centering 
	\includegraphics[width=0.95\columnwidth]{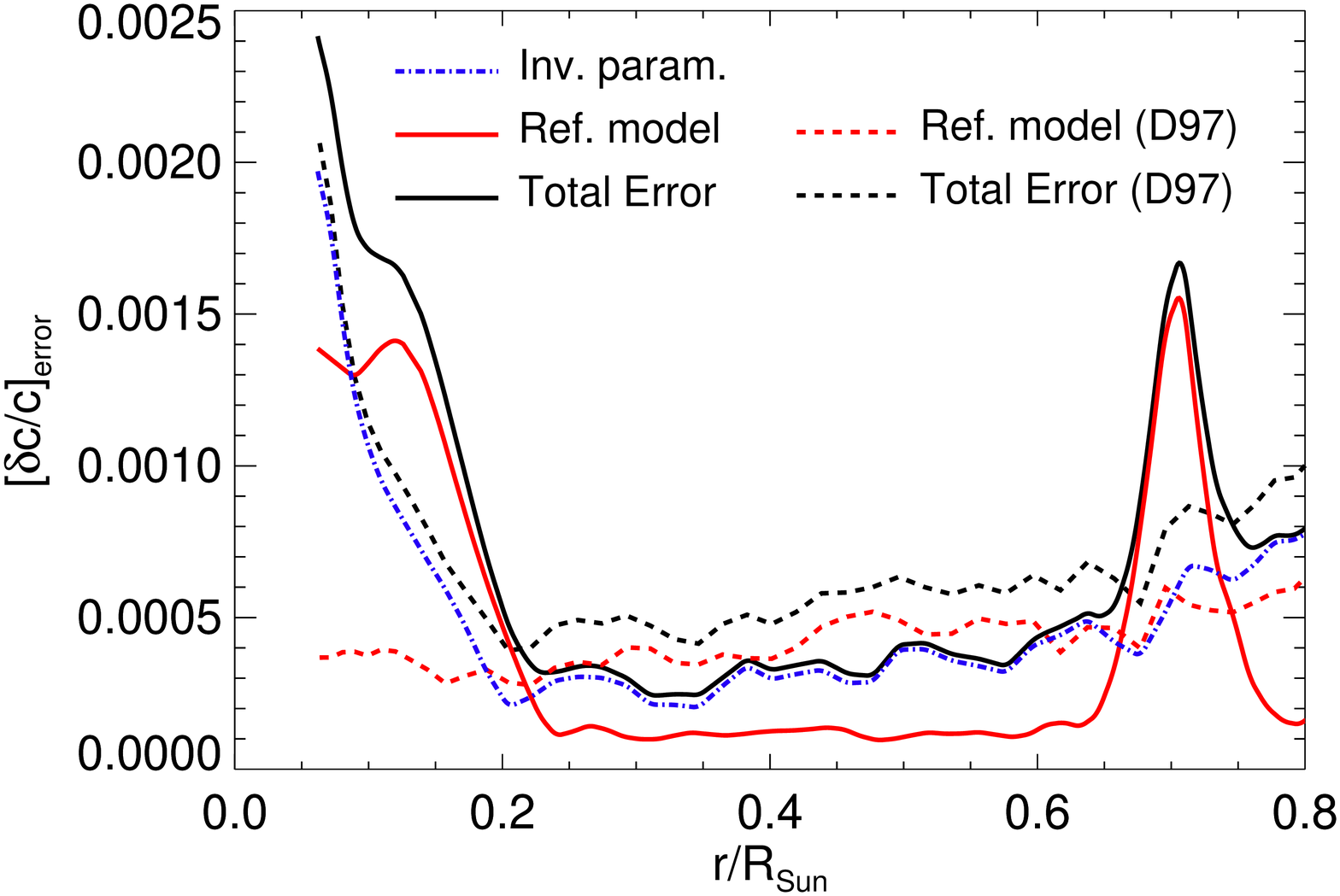}
        \caption{Relevant sources of error for solar sound speed inversions. Solid and dashed lines show respectively our new and the older \citep{deglinnoccenti:1997} estimates of total (black) and reference model (red) errors. The dashes-dotted line shows the error associated to numerical parameters in the inversion procedure that we continue to adopt from \citep{deglinnoccenti:1997}. \label{fig:dcerror}}
\end{figure}

The total error due to the three error sources linked to  $\delta c/c$ inversion is shown in Fig.\,\ref{fig:ssmcs} as the grey shaded area around 0. We have improved the calculation of two of these error sources in  comparison to results in \citet{deglinnoccenti:1997}. The first one is  the error in $\delta c /c$ resulting from propagating the errors  in the observed frequencies. This is now done on the basis of the BiSON-13 dataset, a much more modern dataset with smaller frequency errors. This is not a dominant error source at any location in the Sun. 
More importantly, however, is the dependence of the solar sound speed on the reference model employed for the inversion. Previously \citep{deglinnoccenti:1997,basu:2000}, this dependence was estimated by performing sound speed inversions for a few solar models with different input physics, but with fixed solar composition. Here, instead, we resort to using two sets of 1000 SSMs originally computed by \citet{bahcall06}, with one set based on GS98 and the other one on AGS05 \citep{ags05} solar compositions. In both cases, composition uncertainties used for those datasets correspond to the so-called ``conservative'' uncertainties and are, in fact, about twice as large, or more, as those quoted in the corresponding spectroscopic results. In addition, all other input parameters in SSM calculations have been varied. For these 2000 models, inversions have been carried to determine the solar sound speed profiles. The dispersion of the results, as a function of radius, have been used to derive the dependence of inferred solar sound speed on the inversion reference model. An alternative, and more consistent approach, would be to perform inversions for all the models in our MC simulations, as was done in \citet{bahcall06}. This is a very time consuming procedure because it is not fully automated and we decided not to repeat it in the present paper. But our approach, just described, makes use of a broad range of SSMs and ensures a conservative estimate of this error source. A comparison of our current estimates of uncertainties with respect to previous estimates is shown in Fig.\,\ref{fig:dcerror}, where solid and dashed lines depict currently adopted and older errors respectively.

Using model and inversion uncertainties as described above, we compare how well the predicted sound speed profiles of B16-GS98 and B16-AGSS09met  agree with helioseismic inferences. For this, we use the same 30 radial points employed in \citet{villante14}. We use the models in the MC simulations to obtain the covariance matrix for these 30 points and assume inversion uncertainties at different radii as uncorrelated. We acknowledge the latter is an assumption and we expect to improve on it in the future. Results are shown in the second row of Tab.\,\ref{tab:chi2}. For 30 degrees-of-freedom (dof), B16-GS98 gives $\chi^2= 58$, or a 3.2$\sigma$ agreement with data. For B16-AGSS09met results are $\chi^2=76.1$, or 4.5$\sigma$. Below we analyze in some detail the significance of these results.

It is apparent from Fig.\,\ref{fig:ssmcs} that, at almost all radii, the sound speed profile of  B16-GS98  fits well within the 1$\sigma$ uncertainties. This is true even for the peak right below the CZ at $r/\rsun \approx 0.6-0.7$. But looks can be deceiving. The difference between B16-GS98  and the Sun is dominated by wiggles of relatively small amplitude. However, changes in input quantities, including radiative opacities, do not lead to  variations in SSM sound speeds on small radial scales, so values of the sound speed at different radii in solar models are strongly correlated. Including these correlations by means of a covariance matrix in the calculation of $\chi^2$ explains why the large value $\chi^2=58$ is obtained for the B16-GS98 which, apparently, fits well within 1$\sigma$ contours. This results reflects the fact that, within the framework of SSMs and our treatment of uncertainties, particularly of opacities, it is not possible to find a combination of input parameters that would make the wiggles go away. To confirm this, we checked that when the covariance matrix is assumed diagonal, i.e. correlations are neglected,  $\chi^2=23.6$ for the sound speed profile of B16-GS98, well within a 1$\sigma$ result as expected by a naive look at Fig.\,\ref{fig:ssmcs}. 

For B16-AGSS09met, the discrepancy with the solar sound speed is dominated by the  large and broad peak in $0.35 < r/\rsun < 0.72$. In this case, correlations in the model sound speed decrease the level of disagreement with the data. Variations in the model leading to improvements in the sound speed profile will do so at a global scale. If, as a test, sound speed correlations are neglected for this model, we obtain a larger $\chi^2=100.4$, i.e. the opposite behavior than for B16-GS98. 

It is important to notice that in the case of B16-GS98, the largest contribution to the sound speed $\chi^2$ comes from the narrow region $0.65 < r/\rsun < 0.70$ that comprises 2 out of all the 30 points. If these two points are removed from the analysis $\chi^2$ is reduced from 58 to 34.7, equivalent to a 1.4$\sigma$ agreement with the solar sound speed (entry identified as $\delta c/c$ no-peak in Table~\ref{tab:chi2}). For B16-AGSS09met this test leads to a 2.7$\sigma$ result. This exercise highlights the qualitative difference between SSMs with different compositions; it shows that for GS98 the problem is highly localized whereas for  AGSS09met  the disagreement between SSMs and solar data occurs at a global scale, i.e. the \emph{solar abundance problem}. 

The result of removing two points from the sound speed analysis is reassuring in that the GS98 composition leads to a SSMs that is in quite good agreement with solar data. But it also highlights limitations of SSMs in providing an accurate description of the solar region just below the convective envelope, a fact known for a long time, e.g. \citet{chitre:1998,jcd:2011}.

Table\,\ref{tab:ssmres} lists additional quantities and their model uncertainties associated with the solar calibration: mixing length parameter ($\alpha_{\rm MLT}$), initial helium $\yini$ and metallicity $\zini$ and the corresponding surface ($\ysur$ and $\zsur$) and central $\ycen$ and $\zcen$ quantities, $\rcz$ as stated above and the average rms difference of the relative sound speed difference shown in Fig.\,\ref{fig:ssmcs}. As discussed in this section, this rms value is only indicative of the quality of the models because it neglects correlations in the models.

\subsection{Neutrino fluxes}\label{sec:neutrinos}

The most relevant updates in the B16 SSMs are related to updates in several key nuclear reaction rates (see Sect.\,\ref{sec:updates}) that have a direct effect on the predicted solar neutrino fluxes. The detailed results for all the neutrino fluxes are summarized in Table~\ref{tab:neutrinos}. 

\begin{table}[h]
\setlength{\tabcolsep}{1pt}
\begin{tabular}{c| c c c c}
Flux & B16-GS98 & B16-AGSS09met & Solar$^a$& Chg.\\
\hline
$\phipp$ & $5.98(1 \pm 0.006)$ & $6.03(1 \pm 0.005) $&$5.97^{(1+0.006)}_{(1-0.005)}$& 0.0\\
$\phipep$  &$ 1.44(1 \pm 0.01) $&$1.46(1 \pm 0.009) $&$1.45^{(1+0.009)}_{(1-0.009)}$& 0.0\\
$\phihep$ & $7.98(1 \pm 0.30) $&$8.25(1 \pm 0.30) $&$19^{(1+0.63)}_{(1-0.47)}$& -0.7\\
$\phibe$ &$ 4.93(1 \pm 0.06)$ &$4.50(1 \pm 0.06) $&$4.80^{(1+0.050)}_{(1-0.046)}$& -1.4 \\
$\phib$ & $5.46(1 \pm 0.12)$ &$4.50(1  \pm 0.12) $&$5.16^{(1+0.025)}_{(1-0.017)}$& -2.2\\
$\phin$ & $2.78(1 \pm 0.15)$ &$2.04(1  \pm  0.14) $&$\le 13.7$ & -6.1\\
$\phio$ & $2.05(1 \pm 0.17)$ &$1.44(1 \pm 0.16) $&$\le 2.8$& -8.1\\
$\phif$ & $5.29(1 \pm 0.20)$ &$3.26(1 \pm 0.18) $&$\le 85$& -4.2\\ \hline
\end{tabular}
\caption{Model and solar neutrino fluxes. Units are: $10^{10}$\,(pp), $10^{9}\,{\rm(^7 Be)}$, $10^8\,\rm{(pep,\,^{13}N,\,^{15}O})$, $10^6\,(\rm{^8B,^{17}F)}$ and $10^3\rm{(hep)}$ $\rm{cm^{-2} s^{-1}}$. $^a$Solar values from \citet{bergstrom16}. Last column corresponds to the relative changes (in \%) with respect to SSMs based on SFII nuclear rates, which are almost independent of the reference composition.}
\label{tab:neutrinos}
\end{table}

Currently, $\phib$ and $\phibe$  are the fluxes most precisely determined experimentally, and can be used to perform a simple test of the models. Furthermore, these are also the two fluxes from the pp-chains that are most sensitive to temperature, i.e. to the conditions in the solar core and the inputs in solar models. In \citet{serenelli11}, the agreement between SFII-GS98 and SFII-AGSS09met  with solar fluxes was virtually the same. The solar values determined from experimental data in that work were $\phib= 5 \times 10^6\,\hbox{cm$^{-2}$s$^{-1}$}$ and $\phibe=4.82 \times 10^9\,\hbox{cm$^{-2}$s$^{-1}$}$ with 3\% and 4.5\% uncertainties respectively. SFII-GS98 yields $\phib= 5.58 (1\pm0.14)\times 10^6\,\hbox{cm$^{-2}$s$^{-1}$}$  and $\phibe=5.00(1\pm0.07) \times 10^9\,\hbox{cm$^{-2}$s$^{-1}$}$, while SFII-AGSS09met  gives $\phib=4.59\times 10^6\,\hbox{cm$^{-2}$s$^{-1}$}$ and $\phibe=4.56 \times 10^9\,\hbox{cm$^{-2}$s$^{-1}$}$ with same fractional errors as SFII-GS98. Experimental results for both $\phib$ and $\phibe$ were right in between the predictions for the two SSMs.

\begin{figure}
	\centering
	\includegraphics[width=0.95\columnwidth]{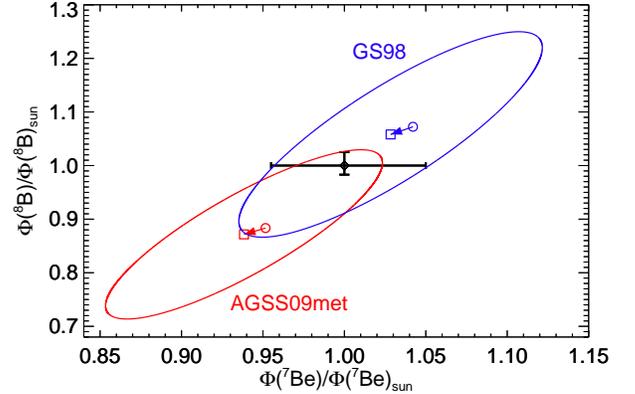}
        \caption{$\phib$ and $\phibe$ fluxes normalized to solar values \citep{bergstrom16}. Black circle and error bars: solar values. Squares and circles: results for B16 (current) and SFII (older) generation of SSMs respectively. Ellipses denote theoretical 1$\sigma$ C.L. for 2 dof. \label{fig:arrows}}
\end{figure}

The new B16 generation of solar models, together with the recent determination of solar fluxes by \citet{bergstrom16} included in Table~\ref{tab:neutrinos}, leads to some changes in this stalemate. On one hand, SSM predictions for $\phib$ and $\phibe$ are reduced for both GS98 and AGSS09met compositions by about 2\% with respect to previous SFII SSM due to  the larger ${\rm p(p,e^+\nu_e)d}$ rate (for $\phib$ this is partially compensated  by the increase in the ${\rm ^7Be(p,\gamma)^8B}$ rate). On the other hand, solar fluxes determined by \citet{bergstrom16} results in a solar $\phib$ that is about 3\% higher than the value used in \citet{serenelli11}. \citet{bergstrom16} includes in their analysis the Borexino Phase-2 data \citep{borexino:2014}, the combined analysis of all three SNO phases \citep{sno:2013} and results from phase IV of Super-Kamiokande \citep{superkIV:2016} all of which were not available in 2011.  Figure~\ref{fig:arrows} reflects the updated state of matters, and shows model and experimental results normalized to the newly determined solar values. Central values of B16-GS98 are closer to the solar values, both for $\phib$ and $\phibe$, than B16-AGSS09met fluxes. Both solar compositions lead to SSMs, however, that are consistent with solar neutrino fluxes within 1$\sigma$. A comparison between models and solar data for these two fluxes yields $\chi^2({\rm GS98})= 0.2$ and $\chi^2({\rm AGSS09met})= 1.45$ (see Table~\ref{tab:chi2}). This calculation includes model correlations obtained from the MC simulations and the distribution of solar fluxes from \citet{bergstrom16}.  

As discussed in Sect.\,\ref{sec:updates}, recent determinations of $S_{34}(0)$ range between $5.42\times10^{-4} \,\rm{MeV\,b}$ to $5.72\times 10^{-4}\,\rm{MeV\,b}$. We speculate here on the impact of adopting the slightly lower $S_{34}(0)$ value such as determined by \citet{deboer14} (see Sect.\,\ref{sec:updates}). A 3.2\% reduction in $S_{34}(0)$ leads to a decrease in $\phib$ and $\phibe$ of about 2.7\% and 2.8\% respectively. This change leads to $\chi^2({\rm GS98})= 0.13$ (0.1$\sigma$) and $\chi^2({\rm AGSS09met})= 2.4$ (1.0$\sigma$). In this hypothetical scenario agreement between  B16-AGSS09met  and data is slightly larger than 1$\sigma$. Although this would still be far from being too useful as a discrimination test between solar models, this exercise helps in showing that the few percent systematics present in the determination of nuclear reaction rates can still be a relevant source of difficulty in using neutrino fluxes as constraints to solar model properties. 

The most important changes in the neutrino fluxes occur for $\phin$ and $\phio$, in the CN-cycle. These fluxes are potentially excellent diagnostics of properties of the solar core. In particular, their dependence on the metallicty is two-fold: through opacities much in the same manner as pp-chain fluxes, and also through the influence of the added C+N abundance in the solar core. This latter dependence makes these fluxes a unique probe of  the metal composition of the solar core. The expectation values in the B16 SSMs are about 6\% and 8\% lower than for the previous SFII models for $\phin$ and $\phio$ respectively. This results from the combined changes in the p+p and $^{14}$N+p reaction rates (Table~\ref{tab:params1}). 

CN fluxes have not yet been determined experimentally. The global analysis of solar neutrino data performed by \citet{bergstrom16} yields the upper limits that we include in Table~\ref{tab:neutrinos}. The Borexino collaboration, based on a different analysis of Borexino data alone, has reported an upper limit for the added fluxes $\phin + \phio < 7.7\times 10^8 \hbox{cm$^{-2}$s$^{-1}$}$ \citep{borexino:2012}.

We close this section with a comparison of models and solar data for all neutrino fluxes. $\chi^2$ values are 6.01 and 7.05 for B16-GS98 and B16-AGSS09met models respectively and are also included in Tab.\,\ref{tab:chi2}. This global comparison is clearly dominated by the $\phib$ and $\phibe$ fluxes. It is evident that current determination of solar neutrino fluxes are well described by models with any of the two solar compositions. 

\subsection{Global analysis}

What is the performance of both B16 SSMs when all the observables discussed before are used for the comparison? Results are summarized in the last two rows of Tab.\,\ref{tab:chi2} when all the sound speed profile is used or the two points in the region $0.65 < r/\rsun < 0.70$ are excluded. Global $\chi^2$ is not strictly the sum of the individual contributions because of correlations between, e.g. $\rcz$ and the sound speed profile. Deviations are however small. 

Final $\chi^2$ values are dominated by the sound speed for both models, although $\ysur$ and $\rcz$ are also relevant for B16-AGSS09met. The global analysis yields a not too good 2.7$\sigma$ result for B16-GS98. However, this is strongly linked to the behaviour of the sound speed profile right below the convective zone, as explained in Sect.\,\ref{sec:helio}. Excluding two points in the sound speed lead to an overall, comforting, 0.9$\sigma$ agreement of this model with solar data. In the case of B16-AGSS09met, the overall agreement with the data is quite poor, at 4.7$\sigma$, which improves to only 3.0$\sigma$ if the critical points in the sound speed profile are excluded. This is still a poor agreement with data. 

It is interesting here to consider the impact of radiative opacity in the results we obtain. We have assumed a 7\% uncertainty for the opacity at the base of the convective zone. Different authors have estimated that changes between 15 and 20\% at that location are required to solve the solar abundance problem. Therefore, it may seem somewhat surprising that AGSS09met yields a much larger, 4.7$\sigma$, disagreement. Naively, we would expect a disagreement at approximately 3$\sigma$ or smaller level provided the level of uncertainty we adopt for opacity. It should be noted, however, that the linear behavior for the opacity error function (Eq.\,\ref{eq:dkappaerror}) permits to compensate the differences between the AGSS09met and GS98 SSMs but it is not flexible enough (for both compositions) to accommodate a better fitting sound speed profile.

 A more detailed analysis of the error function of the opacity requires a more flexible implementation. Because the shape of the error function is itself unknown, a non-parametric approach, where no a priori assumptions are made about this shape, is a promising avenue to treat this problem. This is beyond the scope of this paper, but it is work in progress and will be reported in a forthcoming publication (Song et al. in preparation).

\section{Errors in standard solar models}
\label{sec:ssmerrors}

Model uncertainties for all solar quantities $Q$ of interest are listed in Tables~\ref{tab:ssmres} and \ref{tab:neutrinos} and correspond to 68\% C.L.

We base our model error determination in a MC approach, as introduced in \citet{bahcall88} and later used in \citet{bahcall06}. This is described in Sect.\,\ref{sec:MonteCarlo}. The use of MC simulations has the advantage that it does not require the assumption of linear response of the models to changes in the input parameters. Linearity of solar models is usually a very good approximation because the uncertainties of most of the input parameters are small. However, deviations from a linear response are seen when $2\sigma$ variations of diffusion coefficients and abundances of elements such as C, N and O are considered. MC calculations ensures, then, a consistent assessment of errors. It is true, however, that MC hardwires uncertainties and makes it more difficult to single out relevant individual uncertainty sources for specific observables. Here, power-law expansions are very useful and we in fact resort to them to identify dominant sources of uncertainty for neutrino fluxes and helioseismic probes (Sect.\,\ref{sec:powerlaws}). We close the discussion on errors by considering in some detail the impact of our specific choice of opacity uncertainty (Eq.\,\ref{eq:dkappaerror}) in comparison to the previously adopted error measure (OP-OPAL difference; \citealp{bahcall06,villante14}).

\begin{figure*}
\centering 
\includegraphics[width=0.32\textwidth]{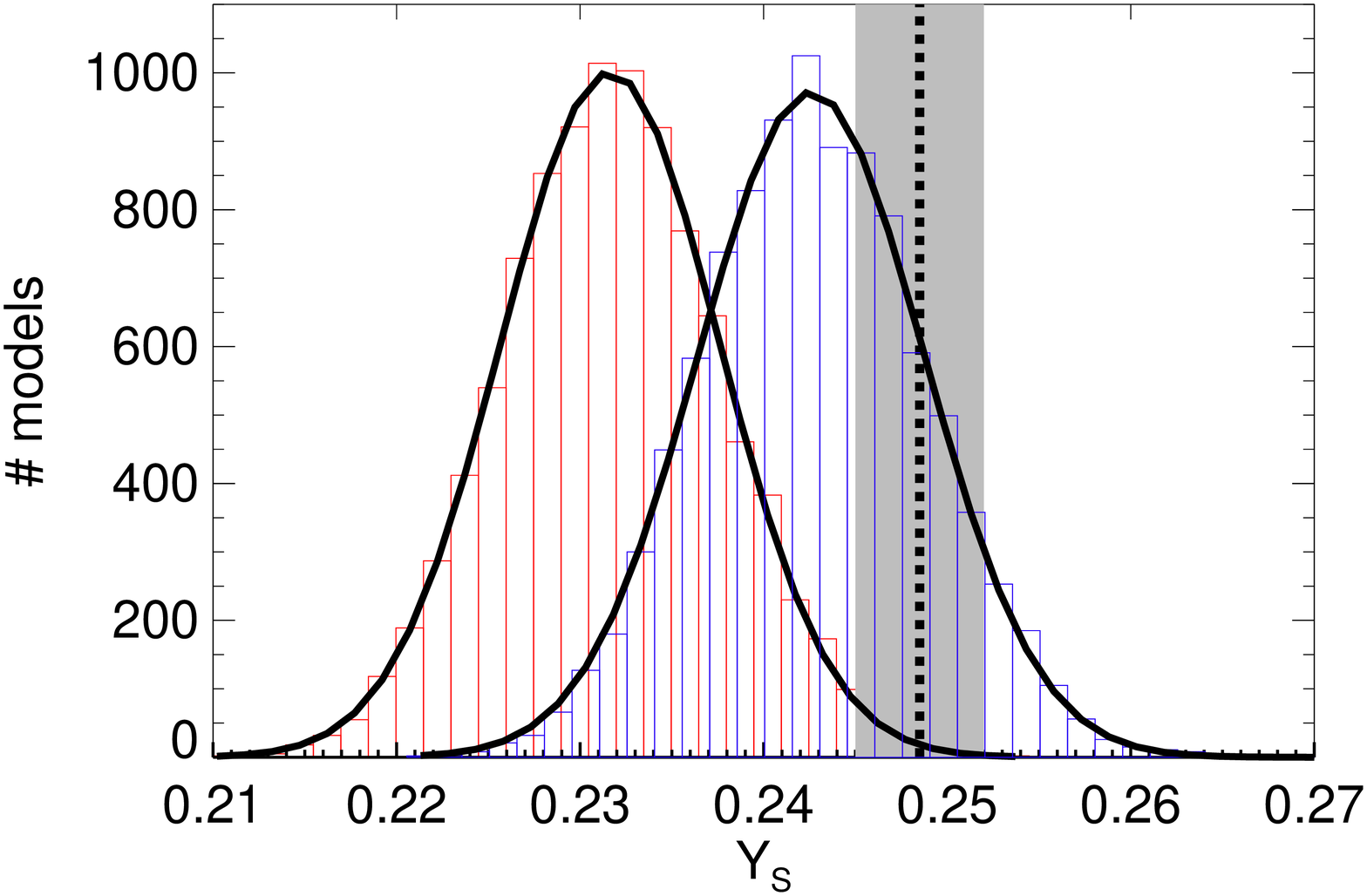}
\includegraphics[width=0.32\textwidth]{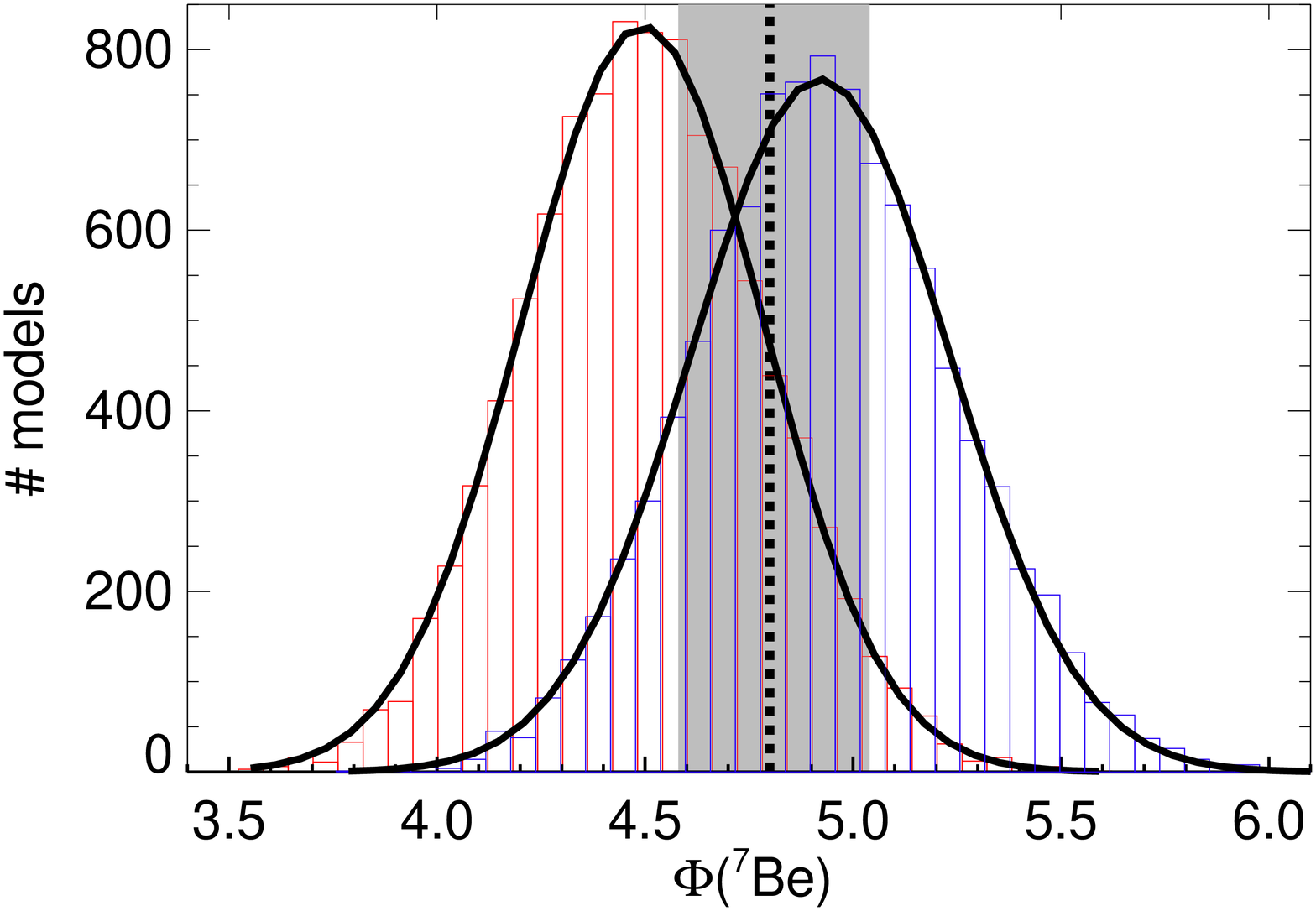}
\includegraphics[width=0.32\textwidth]{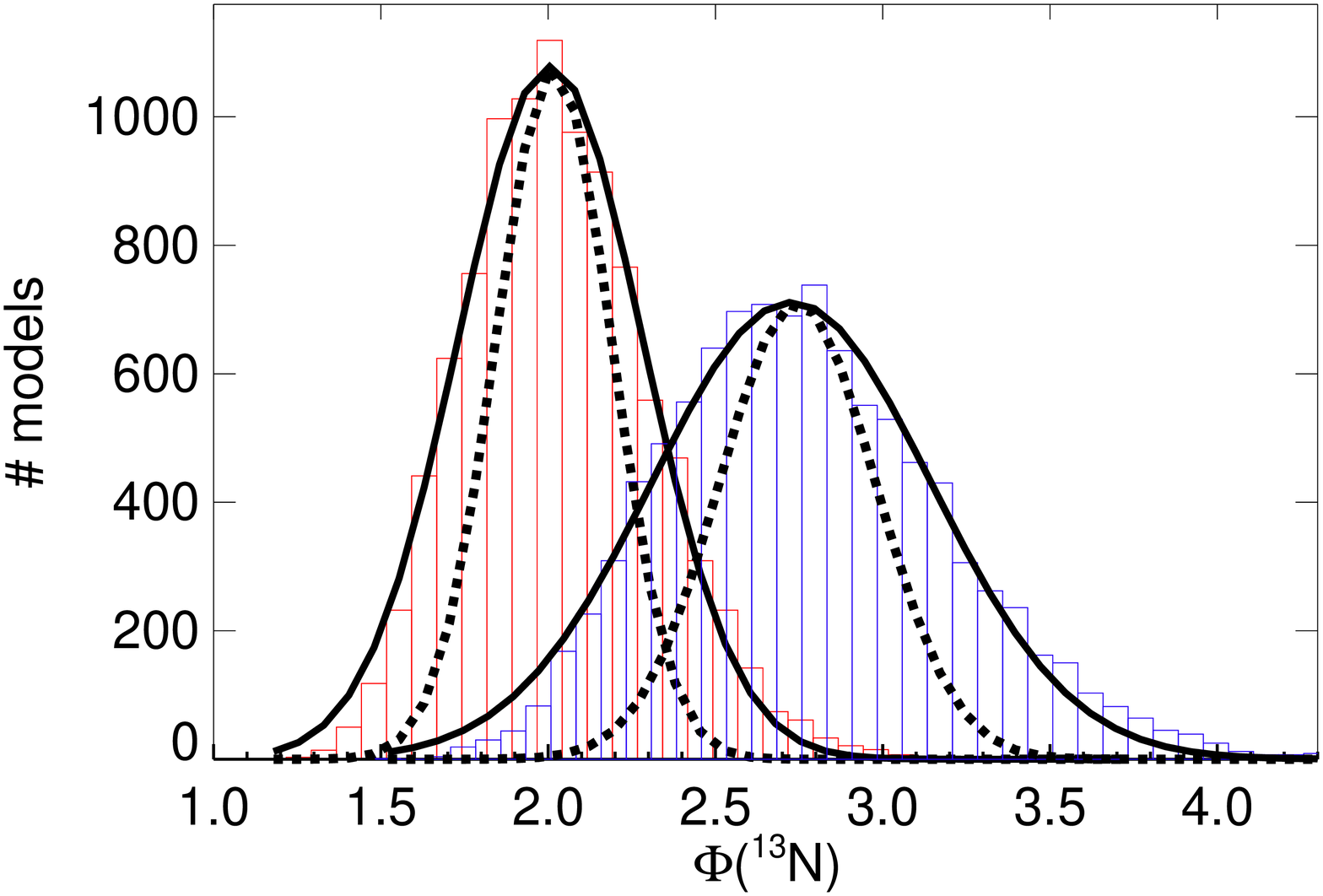} \\
\includegraphics[width=0.32\textwidth]{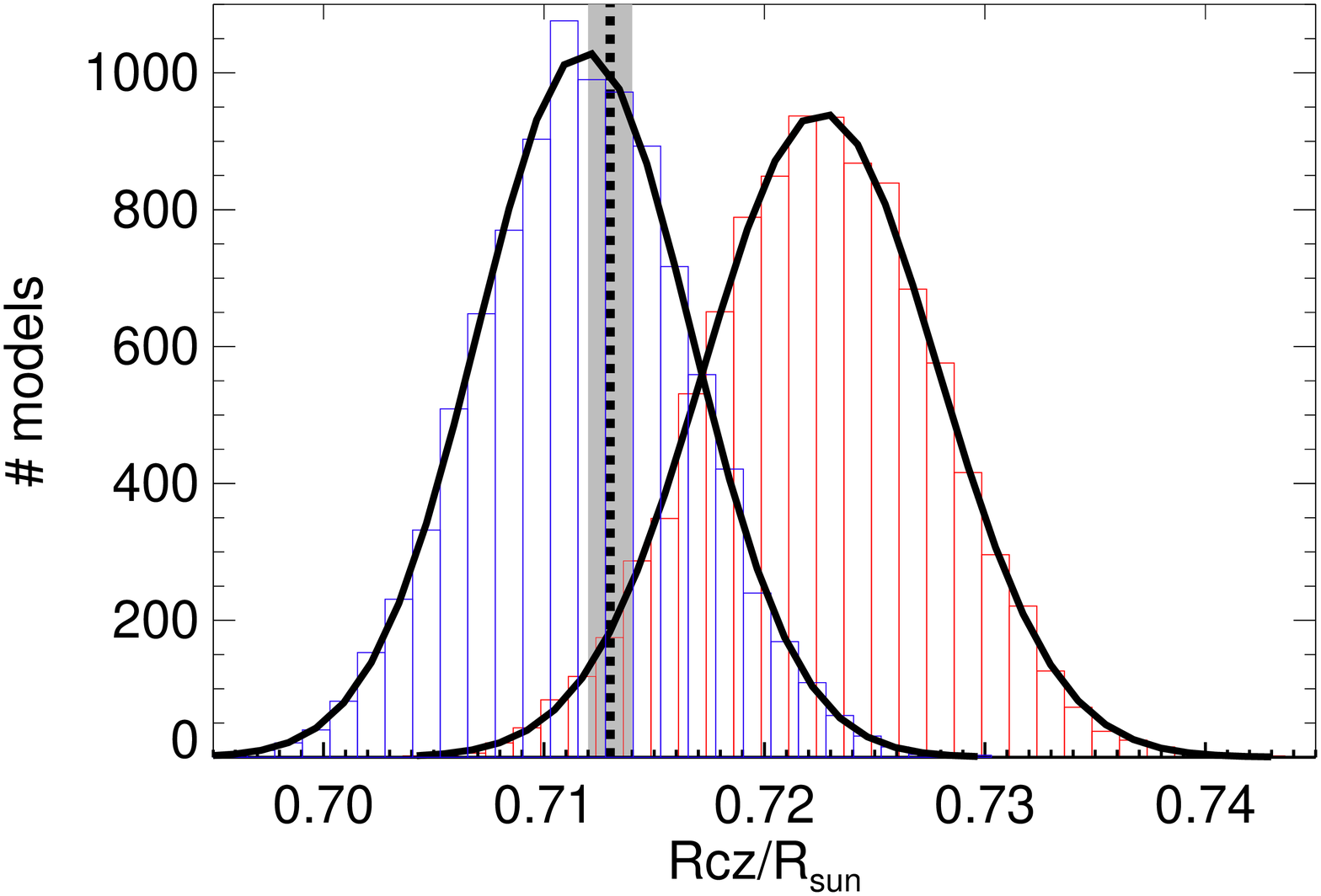}
\includegraphics[width=0.32\textwidth]{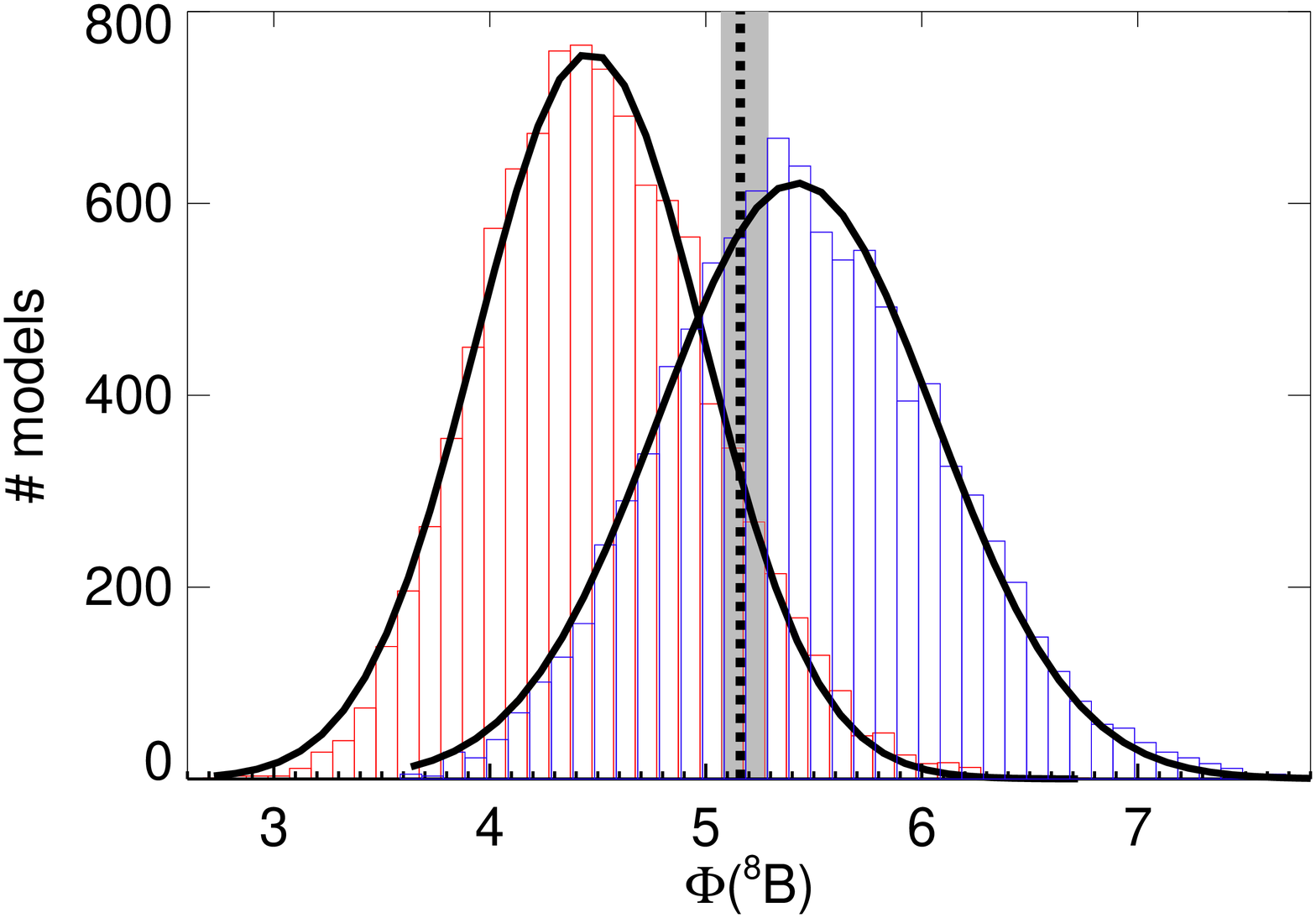}
\includegraphics[width=0.32\textwidth]{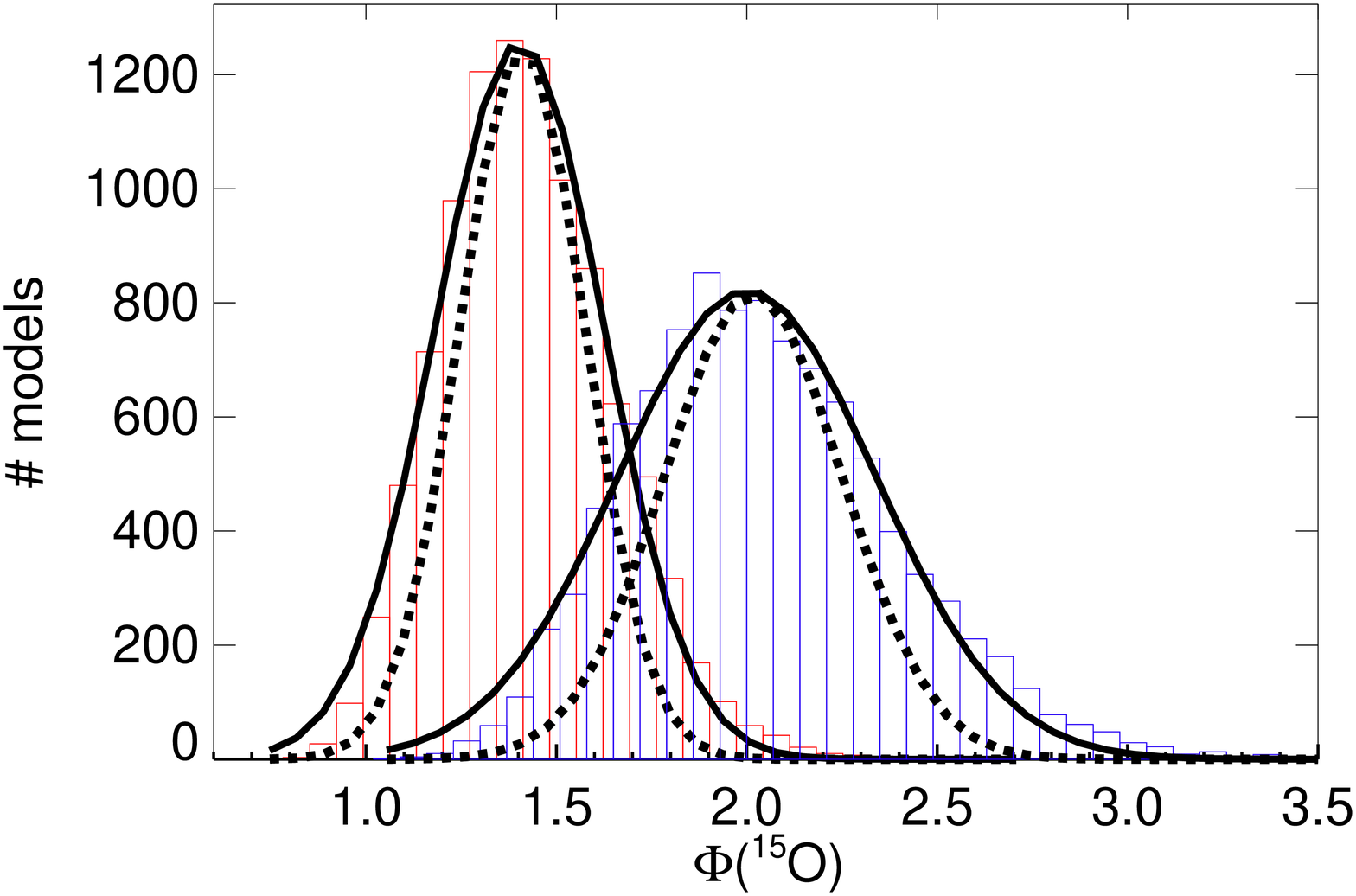}
\caption{Monte-Carlo results for neutrino fluxes and main properties of SSMs. Blue shows results for GS98 and red for AGSS09met. Black solid lines shows the associated Gaussian distributions and the dashed lines, for $\phin$ and $\phio$, the distributions neglecting errors from solar composition. Vertical black dotted lines show the observational results and the grey shaded region the associated errors. Units for neutrino fluxes are as in Table \ref{tab:neutrinos}.\label{fig:mc_histo2}}
\end{figure*}

\subsection{Monte Carlo simulations}\label{sec:MonteCarlo}

We construct MC simulations of SSMs following the procedure used in \citet{bahcall06}. A large set of 10000 SSMs is constructed where, for each model, the values of the input quantities $\{I\}$ are chosen randomly from their respective distributions. Here, $\{I\}$ is the set of input parameters including composition of chemical elements, diffusion rate, $\lsun$, $\tausun$ and nuclear cross section parameters. They are treated as follows.

\subsubsection{Treatment of uncertainties}

\paragraph{Composition parameters} 

Two different MC sets are computed, one per reference composition, i.e. GS98 and AGSS09met (Table\,\ref{tab:compo}). In each set, the $j$-th SSM is calculated by assuming:
\begin{equation}
\epsilon_{i,j} = \overline{\epsilon}_i + C_{i,j} \cdot \sigma_i,
\end{equation}
where $\overline{\epsilon}_i$  and $\sigma_i$ are the central value and error for  each element $i$ (different for each reference composition) and the factors $C_{i,j}$ are sampled from independent univariate gaussian random distributions. Note $\overline{\epsilon}_i$ is a logarithmic measure of abundance.

\paragraph{Other input parameters}
In Tables~\ref{tab:params1} and \ref{tab:params2} we give the central values $\overline{I}$ and the fractional uncertainties $\sigma_{I}$ 
of the 11 non-composition input parameters that are varied in our MC runs. The $j$-th SSM is calculated by assuming
\begin{equation}
I_j = \overline{I}\left(1+ A_{I,j} \cdot \sigma_I\right),
\label{eq:mc_distr1}
\end{equation}
where the factors $A_{I,j}$ are sampled from independent univariate gaussian random distributions. For the case of diffusion, the central value ``1'' in Table~\ref{tab:params2} refers to the standard coefficients used in \texttt{GARSTEC}, computed following the method of \citet{thoul94}.

\paragraph{Opacities}
As described in Sect.\,\ref{sec:uncertainties}, the contribution of opacity uncertainty to all quantities $Q$ are included \emph{a posteriori} by means of Eq.\,\ref{eq:dkappaerror}. The linear response of solar models to opacity variations ensures that this procedure is sufficiently accurate. In return, our set of MC calculations can be used to test other opacity error functions, e.g. the OP-OPAL difference (Sect.\,\ref{sec:opacityerrors}) or others motivated by future theoretical or experimental work on opacities.

With the opacity kernels in hand our implementation of opacity uncertainties is very simple. After the $j$-th SSM is calculated, a change of the opacity profile $\delta \kappa(T)$ is modelled using Eq.\,\ref{eq:dkappaerror}, with the coefficients $a_j$ and $b_j$ being extracted from independent gaussian distributions with zero means and dispersions $\sigma_a=2\%$ and $\sigma_b=6.7\%$, respectively, values that reflect our estimates of the magnitude of opacity uncertainties (Sect.\,\ref{sec:uncertainties}). The effect of the opacity variation on the various SSM predictions $Q$ is then calculated by using the kernels $K_{Q}(T)$ and the estimated change $\delta Q$ then added to the SSM prediction.

\subsubsection{Monte-Carlo Results} 

The distributions of important quantities resulting from the MC simulations are presented in Fig.\,\ref{fig:mc_histo2} for GS98 and AGSS09met compositions. We have used these distributions to compute the uncertainties $\sigma_Q$ (68.3\% C.L.) for model predictions of all quantities $Q$ given in this work, in particular results reported in Tables \ref{tab:ssmres} and \ref{tab:neutrinos}. Plots for $\ysur$, $\rcz$, $\phibe$, and $\phib$ show the solar values as determined from helioseismology and solar neutrino experiments.

Overall, distributions are well described by Gaussian distributions, also included in Fig.\,\ref{fig:mc_histo2} as thick solid lines. The distributions are slightly skewed only for $\phin$ and $\phio$, with a longer tail towards higher values. This is because for chemical elements the distribution is assumed Gaussian for their logarithmic abundance, and these fluxes depend linearly on the added C+N abundance. Then, at least formally, the resulting distributions for these fluxes are described better by log-normal functions. Deviations are small, however, and we prefer to ignore them by quoting symmetric uncertainties (Table~\ref{tab:neutrinos}). For these neutrino fluxes we also show the Gaussian distribution resulting when composition errors are neglected. These uncertainties represent the straw man separation power CN fluxes have for discriminating between different sets of solar composition if no other information is used. This can be largely improved, however, using the method developed in \citet{haxton:2008} and \citet{serenelli13}, in which an appropriate ratio of $\phib$ and the CN fluxes is used to cancel out the effect of environmental uncertainties that affect similarly the $\phib$ and CN fluxes in solar models. 

The sound speed difference profile for each of the MC models has been obtained as $\delta c/c = (c_{\odot,C}-c_{\rm mod})/c_{\rm mod}$. Here $c_{\odot,C}$ is the solar sound speed profile and $C$ identifies the composition of the reference SSM used in the inversion. Specifically, we use either B16-GS98 or B16-AGSS09met depending on whether the  MC model belongs to the GS98 or AGSS09met set. This is not a self-consistent procedure because the inferred solar sound speed has a formal dependence on the reference model used for the inversion. This  source of uncertainty then has to be taken into account in an explicit manner, as it has been described in \S~\ref{sec:helio}.

\subsection{Dominant sources of errors} 
\label{sec:powerlaws}

In order to estimate the different contributions $\delta Q_{I}$ to the total error of the quantity $Q$, 
we follow the standard approach and calculate
\begin{equation}
\delta Q_{I} = \alpha_{Q,I} \, \sigma_I
\label{eq:dQI}
\end{equation}
where $\sigma_I$ is the $1\sigma$ fractional uncertainty of the $I$ input parameter and $\alpha_{Q,I}$ is defined by:
\begin{equation}
\alpha_{Q,I} \equiv \frac{\partial \ln Q}{\partial \ln I }
\label{eq:powerlaw}
\end{equation}
The log-derivatives $\alpha_{Q,I}$ are calculated numerically by varying the various input parameters, one at a time, over a range typically larger than their respective 3$\sigma$ uncertainty. The values obtained using the new SSM calculations are similar but update previous determinations \citep{serenelli13}.

The opacity error is described in terms of the two independent parameters $a$ and $b$, equation~\ref{eq:dkappaerror}, that fix the scale and the tilt of the opacity profile.  The derivative of a given quantity $Q$ with respect to these parameters can be calculated from equation~\ref{eq:dqkappa} as:
\begin{eqnarray}
\nonumber
\alpha_{Q,a} &\equiv& \frac{\partial \ln Q}{\partial a } = \int \frac{dT}{T} K_{Q} (T) \\
\alpha_{Q,b} &\equiv& \frac{\partial \ln Q}{\partial b } = \int \frac{dT}{T} K_{Q} (T) \frac{\log_{10}(T/T_{\rm 0})} {\Delta}
\end{eqnarray}
The total error due to opacity is estimated by combining the two contributions in quadrature, i.e.:
\begin{equation}
\delta Q_{\kappa} = \sqrt{(\alpha_{Q,a} \, \sigma_a)^2 + (\alpha_{Q,b} \, \sigma_b)^2} \label{eq:dQopa}
\end{equation}

Results obtained with equations (\ref{eq:dQI}) and (\ref{eq:dQopa}) are presented for the dominant error sources in Tab.\,\ref{tab:quantity}  for all relevant solar quantities. Dominant error sources can be roughly grouped as: composition, nuclear, and stellar physics, the latter dominated by opacity and microscopic diffusion. 

\paragraph{Composition} Errors from composition are dominated by the C, O and Ne. This is not related to the solar composition problem, however, but just to the fact that even the most optimistic spectroscopic determinations of solar abundances have a level of uncertainty of about 10-12\% that is very difficult to beat. Refractories, on the other hand, are more precisely measured from meteorites so their contribution to uncertainties in solar quantities is currently minimal. Clearly, CNO neutrino fluxes are directly affected by these uncertainties which are, in fact, the dominant error sources. For the same reason, $\zini$ (and $\zsur$) error is also dominated by uncertainties in these elements. For helioseismic quantities, O affects $\rcz$ because it is a dominant contributor to opacity at the base of the convective envelope. On the other hand, $\yini$ and $\ysur$ depend more strongly on Ne due to a combination of its large abundance, impact on opacity at deeper layers and larger error.

\paragraph{Nuclear reactions} Nuclear rates are still an important uncertainty source for neutrino fluxes despite big progress in the field. In particular, errors in S$_{34}$ and S$_{17}$ are still comparable or larger than the uncertainties in the experimental  determinations of $\phib$ and $\phibe$. As discussed in Sect.\,\ref{sec:neutrinos}, the ability of solar neutrinos linked to pp-chains to play a significant role in constraining condition in the solar interior depends, although it is not the only factor, on pinning down errors of nuclear reaction rates to just $\sim 2\%$. For CN fluxes, S$_{114}$ is the dominant error source if composition is left aside. Assuming a precise measurement of CN fluxes becomes available in the future, right now S$_{114}$ is the limiting factor in using such measurement as a probe of the solar core C+N abundance \citep{serenelli13}.

\begin{table}
\centering
\begin{tabular}{l | l@{\hskip 0.1cm}r@{\hskip 0.6cm} l@{\hskip 0.1cm}r@{\hskip 0.6cm} l@{\hskip 0.1cm}r@{\hskip 0.6cm} l@{\hskip 0.1cm}r }
Quant.&\multicolumn{8}{c}{Dominant theoretical error sources in \%} \\
\hline
$\phipp$ & $\lsun$:& 0.3 & $S_{34}$: &  0.3 & $\kappa$: & 0.2 & Diff: & 0.2  \\
$\phipep$ & $\kappa$: &0.5 & $\lsun$: & 0.4 & $S_{34}$: & 0.4 & $S_{11}$: & 0.2 \\
$\phihep$& $S_{\rm hep}$: &30.2 & $S_{33}$: & 2.4 & $\kappa$: & 1.1 & Diff: & 0.5\\
$\phibe$& $S_{34}$: &4.1 & $\kappa$: & 3.8 & $S_{33}$: & 2.3 & Diff: & 1.9 \\
$\phib$& $\kappa$: &7.3 & $S_{17}$: & 4.8 & Diff: & 4.0 & $S_{34}$: & 3.9\\
$\phin$& C: &10.0 & $S_{114}$: & 5.4 & Diff: & 4.8 & $\kappa$: & 3.9\\
$\phio$& C: &9.4& $S_{114}$: & 7.9 & Diff: & 5.6 & $\kappa$: & 5.5 \\
$\phif$& O: &12.6 & $S_{116}$: & 8.8 & $\kappa$: & 6.0 & Diff: & 6.0  \\
\hline
$\alphamlt$& O: &1.3 &  Diff: & 1.2 & $\kappa$: & 0.7  & Ne: & 0.7 \\
$\yini$&  $\kappa$: &1.9 & Ne: & 0.5 & Diff: & 0.4 & Ar: & 0.3 \\
$\zini$&  O: &4.7 & C: & 2.0 & Ne: & 1.7 & Diff: & 1.6 \\
$\ysur$& $\kappa$: &2.2 & Diff: & 1.1 & Ne: & 0.6 & O: & 0.3 \\
$\zsur$&  O: &4.8 & C: & 2.0  & Ne: & 1.8 & $\kappa$: & 0.7 \\
$\rcz$&  $\kappa$: &0.6 &  O: & 0.3  & Diff: & 0.3 & Ne: & 0.2 \\
\hline
\end{tabular}
\caption{Dominant theoretical error sources for neutrino fluxes and the main characteristics of the SSM. \label{tab:quantity}}
\end{table}

\paragraph{Opacity and microscopic diffusion} These are the dominant sources of errors not linked to composition or nuclear reactions. For solar neutrinos, our estimate of the contribution of opacity to the total error is similar to previous calculations \citep{serenelli13} despite the different treatment given to opacity errors. In this work we assume a 2\% uncertainty in the center that increases linearly outwards. Because neutrinos are produced in a localized region, our results are not too different from assuming a constant 2.5\% fractional opacity variation, which was the previous choice. Opacity is the dominant error source for $\phib$, and the second one for $\phibe$. For these fluxes, it is important that opacities in the solar core be known to a 1\% level of uncertainty. Current theoretical work shows variations of about 2\% \citep{Krief16} and experimental measurements are notoriously difficult due to the combination of high temperatures and densities involved. 

Opacity is a dominant uncertainty error source for helioseismic quantities, most notably $\rcz$ and $\ysur$, with our new treatment of uncertainties.   A 7\% opacity uncertainty at the base of the convective envelope implies a 0.6\% change in $\rcz$. This is larger than all other uncertainty sources combined and explains the substantially larger error in $\rcz$ given in this work, 0.7\% (Table~\ref{tab:ssmres}), compared to 0.5\% previously determined \citep{bahcall06}. We observe a similar impact on $\ysur$, for which we estimate now a total 2.5\% uncertainty compared to 1.5\% from previous estimations. 

At first glance, the change in uncertainties for $\rcz$ and $\ysur$ might seem not too large but in fact model uncertainties are now substantially larger than helioseismically inferred ones. Moreover, the larger uncertainties lead to a formally better agreement between solar data and SSMs based on AGSS09met composition, which is now placed at about  2.1$\sigma$ level when $\rcz$ and $\ysur$ are considered, whereas before this was closer to 3.5$\sigma$.  

Microscopic diffusion is typically a smaller source of uncertainty than radiative opacities. However, for CN neutrino fluxes its contribution is larger, only after $S_{114}$ and C. The reason is that accumulation of metals in the solar core due to gravitational settling increases CN fluxes both because it leads to a larger opacity in the solar core but also because the increase in the C+N abundance directly affects the efficiency of the CN-cycle.

\subsection{Opacity uncertainties: effect of different parametrizations} \label{sec:opacityerrors}

As discussed before, there is a certain level of arbitrariness in the choice of the error function for opacity. Our standard choice (Sect.\,\ref{sec:uncertainties})  stems from a comparison of available opacity data from both theoretical and experimental sources but also from an attempt of not being too aggressive (optimistic) in our choice.

\begin{table}
\centering
\begin{tabular}{l|cc}
Quant. & $\sigma_{\rm{OP-OPAL}}$ & $\sigma_{\rm out}=7 \%$\\
\hline
$\phipp$&    0.001  & 0.002 \\
$\phipep$&   0.001  & 0.005 \\
$\phihep$&   0.002  & 0.011 \\
$\phibe$&  0.009 &  0.038 \\
$\phib$&   0.021 &  0.074 \\
$\phin$&  0.012 &  0.039 \\
$\phio$&  0.017 &  0.055 \\
$\phif$&  0.019 & 0.061 \\
\hline
$\ysur$  &  0.004 &  0.021 \\
$\zsur$&    0.001 &  0.007 \\
$\rcz$&   0.001 &  0.005 \\ \hline
\end{tabular}
\caption{Fractional error contribution of opacity to the neutrino fluxes and helioseismic quantities SSMs for different choices of opacity variations, as described in the text. \label{tab:qopa}
}
\end{table}

A viable alternative is using the OP-OPAL difference as a 1$\sigma$ measure of the true opacity error \citep{bahcall06,villante10}. Using the opacity kernels, it is straightforward to evaluate the error in solar quantities $Q$ if this opacity error function is used. In Table~\ref{tab:qopa} we compare the fractional errors of solar quantities $Q$ for the linear and the OP-OPAL error functions.  This comparison highlights the enhanced impact of opacity errors on solar quantities following our new approach, which we believe better reflects the current level of uncertainty in stellar opacities. 

\begin{figure}
	\centering 
	\includegraphics[width=0.95\columnwidth]{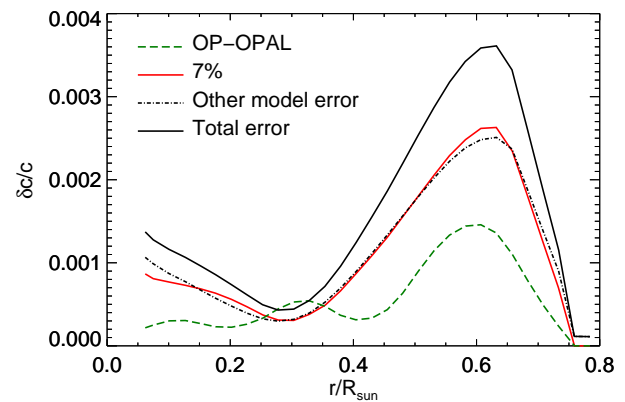}
        \caption{Fractional sound speed variation resulting from different assumptions for the  opacity error (see text). For comparison, the uncertainty due to all other model inputs and the total uncertainty (opacity + other model sources) are also shown. \label{fig:csopa}}
\end{figure}

Finally, Fig.\,\ref{fig:csopa} compares the sound speed uncertainties corresponding to the linear error function (indicated as the 7\% curve in the plot) and the OP-OPAL error function. The former leads to larger uncertainties at almost all radii. Also, the plot includes the total contribution to the sound speed uncertainty from all (model) sources other than radiative opacity. The total uncertainty of the sound speed is also included and it corresponds to the model error budget shown as a pink band in Fig.\,\ref{fig:ssmcs}. The opacity contribution to the total sound speed error matches that of all other uncertainty sources combined, including those from composition errors.

\section{Summary and Conclusions}
\label{sec:summary}

We have presented B16-GS98 and B16-AGSS09met, a new generation of SSMs. They have been computed using the new release of \texttt{GARSTEC} that includes the possibility of using an equation of state consistent with the composition used in the SSM calibration. We have also incorporated the most recent values for the nuclear reactions ${\rm p(p,e^+\nu_e)d}$, ${\rm^3He(^4He,\gamma)^7Be}$ and ${\rm ^{14}N(p,\gamma)^{15}O}$. With respect to previous works, we have implemented a flexible treatment of opacity uncertainties based on opacity kernels that allows testing any arbitrary modification to, or error function of, the radiative opacity profile. Based on current theoretical and experimental results on solar opacities, we have adopted an opacity error that increases linearly from the solar core, where it is 2\%, towards the base of the convective envelope where we assume a 7\% uncertainty. However, we have also tested  the difference between OP and OPAL as the opacity error function. The estimation of central values of solar observables, their uncertainties and model correlations have been obtained from large MC sets of simulations of SSMs that comprise 10000 SSMs per reference composition (either GS98 or AGSS09met). SSMs have been compared against different ensembles of solar observables: $\ysur$ and $\rcz$, sound speed profile, solar neutrinos, and a global comparison that includes the three classes of observables. 

We summarize our most important findings here:

\begin{itemize}

\item Central values for $\phibe$ and $\phib$ in B16 SSMs are reduced by about 2\% with respect to the previous generation of models that were based completely on the A11 nuclear reaction rates. $\phin$ and $\phio$, the CN-cycle fluxes, are reduced by 6\% and 8\% respectively. 

\item Solar neutrino fluxes \citep{bergstrom16} are reproduced almost equally well by both B16-GS98  and B16-AGSS09met,, with only a very minor preference for B16-GS98 ($\chi^2=6.01$ versus 7.05). If only $\phibe$ and $\phib$ are considered, then $\chi^2=0.21$ and 1.50 for B16-GS98 and B16-AGSS09met models respectively. 

\item Helioseismic properties of B16 models are almost unchanged with respect to SFII models. However, our estimation of errors is larger due to our more pessimistic assumption of a 7\% uncertainty in the radiative opacity at the base of the convective envelope. Comparison of models against $\ysur$ and $\rcz$ yields a very good agreement for B16-GS98 (0.5$\sigma$) and a poor one (2.1$\sigma$) for B16-AGSS09met. 

\item We have reevaluated some of the sources of uncertainty associated with solar sound speed inversion. This, together with our new adoption of larger opacity uncertainties lead to B16-GS98 and B16-AGSS09met  to an agreement with data at the level of 3.16 and 4.5$\sigma$ respectively.

\item The seemingly, and surprising, bad performance of the sound speed profile of B16-GS98  is caused almost exclusively by the large sound speed difference in the region $0.65 < r/\rsun < 0.70$. It is well known that the structure of the Sun in this narrow range of radius is not well reproduced by standard models. There is a long list of possibilities suggested to explain this deficit in SSMs: a smoother chemical profile as claimed by \citep{chitre:1998} due to, e.g. turbulent mixing \citep{proffitt:1991,jcd:2007}, a smoother transition between an adiabatic and radiative temperature gradient \citet{jcd:2011} due to overshooting, dynamic effects at the tachocline \citep{brun:2011}, among others. Removing this region from the analysis brings the agreement of B16-GS98 and the solar sound speed to a comforting 1.4$\sigma$. However, this discrepancy cannot be ignored and deserves further work.

\item For B16-AGSS09met, the mismatch with the solar sound speed profile is global. Removing the region $0.65 < r/\rsun < 0.70$ leads to a 2.7$\sigma$ discrepancy with the solar sound speed profile. 

\item The comparison of models with all data yields $\chi^2=65$ for B16-GS98 (40 dof) but only 40.5 (38 dof) when the region $0.65 < r/\rsun < 0.70$ is removed. This is equivalent to a very good 0.9$\sigma$ result. For B16-AGSS09met results are $\chi^2=94$, i.e. 4.7$\sigma$ (or $\chi^2=67$ and 3$\sigma$ without the problematic region). B16-GS98 is a better model than B16-AGSS09met at a statistically significant level.

\item The estimated increase in opacity required to solve the solar abundance problem is 15 to 20\% at the base of the convective envelope. By assuming a 7\% opacity uncertainty in that region, it would be naively expected that B16-AGSS09met is discrepant with solar data at a 2-3$\sigma$ level. The much larger, 4.7$\sigma$ discrepancy is due to the fact that the adopted opacity error function Eq.\,\ref{eq:dkappaerror} allows to compensate the diffferences between AGSS09met and GS98 SSMs but is not flexible enough (for both compositions) to accommodate a better fitting sound speed profile. This is in qualitative agreement with \citet{villante14}.

\item We have identified the dominant uncertainty sources for neutrino fluxes. Radiative opacity is the dominant source among the \emph{stellar physics} quantities, followed by the microscopic diffusion rates. Among nuclear reaction rates, the astrophysical factors $S_{34}$, $S_{17}$ and $S_{114}$ should have their uncertainties reduced to allow more precise tests of solar physics based on solar neutrino experiments. A smaller uncertainty for $S_{114}$ will be crucial in determining the abundance of C+N in the solar core when a precision measurement of the $\phin$ and $\phio$ fluxes becomes available. 

\item For helioseismic quantities, opacity and diffusion are the dominant \emph{stellar} uncertainty sources. Volatile elements, particularly O and Ne, also play an important role.

\end{itemize}

A novel and important aspect of our work is the implementation of opacity kernels to generalize the treatment of opacity uncertainties. This opens up the possibility to use the large sets of models computed as part of our MC simulations to study the impact of any arbitrary modification to the solar opacity profile on all solar observables. But the shape of the profile of opacity uncertainties across the solar interior is unknown. As part of follow up work, we are currently implementing an approach to treat opacity modifications based on a non-parametric Gaussian process  approach that does not assume a specific shape for this error function (Song et al. in preparation). This will allow us to test the best possible solutions that opacity variations can offer not only to the solar abundance problem but also to how well any SSM can fit available solar observables.

\section*{Acknowledgements}

The authors want to thank the anonymous referee whose comments improved the manuscript. We also thank Wick Haxton for enlightening discussions about determinations of nuclear reaction rates. NV and AS are partially supported by grants ESP2014-56003-R and ESP2015-66134-R (MINECO) and SGR14-1458 (Generalitat de Catalunya). SB acknowledges  support from NSF grant AST-1514676.  This work of NS and MCG-G is supported by USA-NSF grants PHY-1316617 and PHY-1620628 and together with JB also by EU Networks FP10 ITN ELUSIVES  (H2020-MSCA-ITN-2015-674896) and INVISIBLES-PLUS (H2020-MSCA-RISE-2015-690575). MCG-G and JB also acknowledge support by MINECO grants 2014-SGR-104, FPA2013-46570, and Maria de Maetzu program grant MDM-2014-0367 of ICCUB. MM is supported by EU Networks FP10 ITN ELUSIVES (H2020-MSCA-ITN-2015-674896) and INVISIBLES-PLUS (H2020-MSCA-RISE-2015-690575), by MINECO grants FPA2012-31880, FPA2012-31880, FPA2015-65929-P MINECO/FEDER, UE and by the ``Severo Ochoa'' program grant SEV-2012-0249 of IFT. C.P.G. is supported by Generalitat Valencia  Prometeo  Grant  II/2014/050,  by  the  Spanish  Grant FPA2014-57816-P  of  MINECO  and  by  EU Networks FP10 ITN ELUSIVES (H2020-MSCAITN-2015-674896) and INVISIBLES-PLUS (H2020-MSCA-RISE-2015-690575).

\appendix

\section{Opacity Kernels}

The calculation of the opacity kernels used to evaluate the
contribution of opacity to the uncertainties of the solar properties 
has been done following the procedure presented in
\cite{Tripathy98}. We summarize it in this Appendix.

First, we assume that in the region of the parameter-space
$(\rho,T,Xi)$ which describes the solar plasma during the Sun
evolution, the variation of the opacity can be approximately described
as a function of the temperature only, i.e. the modified 
opacity table $\kappa(\rho,T,X_i)$ is related to the reference opacity
table $\bar\kappa(\rho,T,X_i)$ by:
\begin{equation}
\kappa(\rho,T,X_i) = 
\left[1+\delta \kappa (T)\right] \; \bar\kappa(\rho,T,X_i)
\end{equation}
where $\delta \kappa(T)$ is an arbitrary function.
If the changes of opacity are small enough, i.e. $\delta \kappa (T)\ll
1$,the model responds linearly to these perturbations and the fractional
variation of a generic observable $Q$ can be expressed as:
\begin{equation}
 \delta Q = \int \frac{dT}{T}K_Q (T) \delta \kappa (T).
 \label{eq:dqkappa}
\end{equation}
In the above equation, the kernel $K_Q(T)$ describes the response of
the considered quantity to changes of the opacity at a given
temperature.

Our goal is to obtain these kernels and to use them in order to study
the effects produced by arbitrary opacity changes. To do so, we need to determine the
response of SSMs when opacity is changed in a thin shell
whose position is determined by the temperature. This narrow
perturbation is ideally represented as delta function:
\begin{equation}
\delta \kappa(T) =  C\delta(\ln{T} - \ln{T_0}).
 \label{eq:deltadirac}
\end{equation}
In numerical calculations, the $\delta$-function is approximated by a
gaussian, i.e.:
\begin{equation}
\delta \kappa(T) =  \frac{C}{\sqrt{2\pi} \sigma} \exp \left[
  -\frac{(\ln{T} - \ln{T_0})^2}{2\sigma^2}\right]\, ,
\label{eq:gaussian}
\end{equation}
where we require $\sigma \ll 1$ so that the opacity variations are
sufficiently localized and $C/\sigma \ll \sqrt{2\pi}$ in order to
avoid non linear effects. We choose $\sigma = 0.03$ and $C = 0.004$.
We have calculated a set of SSMs with opacity perturbations located
in the temperature range $\log_{10}{T_0} = 6.3 - 7.2$, where radiative
transport takes place and thus, the uncertainties of the radiative opacities play an important role. 
Finally, the opacity kernels are calculated by using the outputs of
these models and normalizing the obtained variations $\delta
Q$ according to: 
\begin{equation}
K_Q(T_0) = \frac{\delta Q}{C}
\end{equation}
as it is prescribed by Eq. \ref{eq:dqkappa}.

\begin{figure}
	\centering 
    \includegraphics[width=0.49\columnwidth]{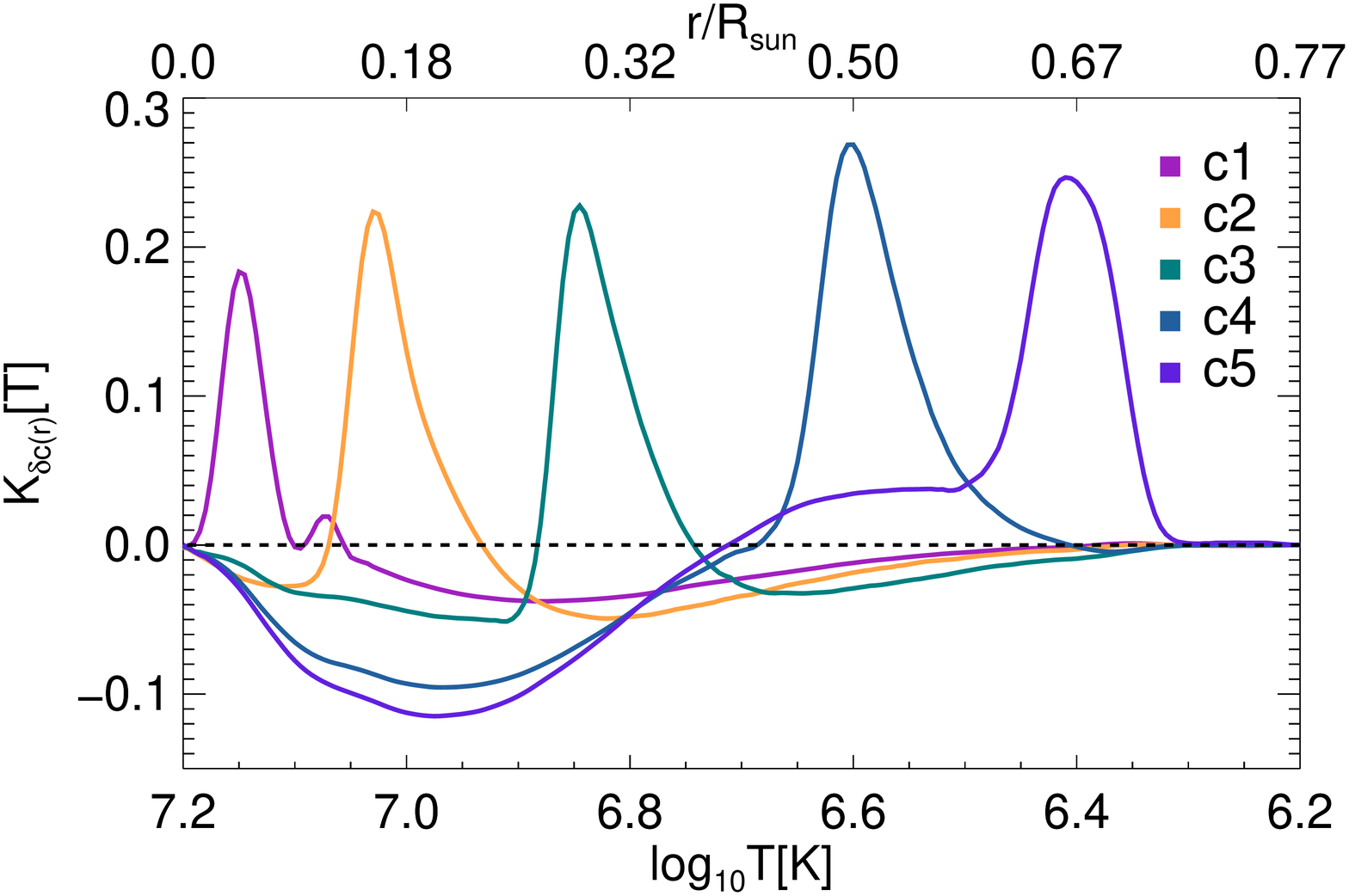}
	\includegraphics[width=0.49\columnwidth]{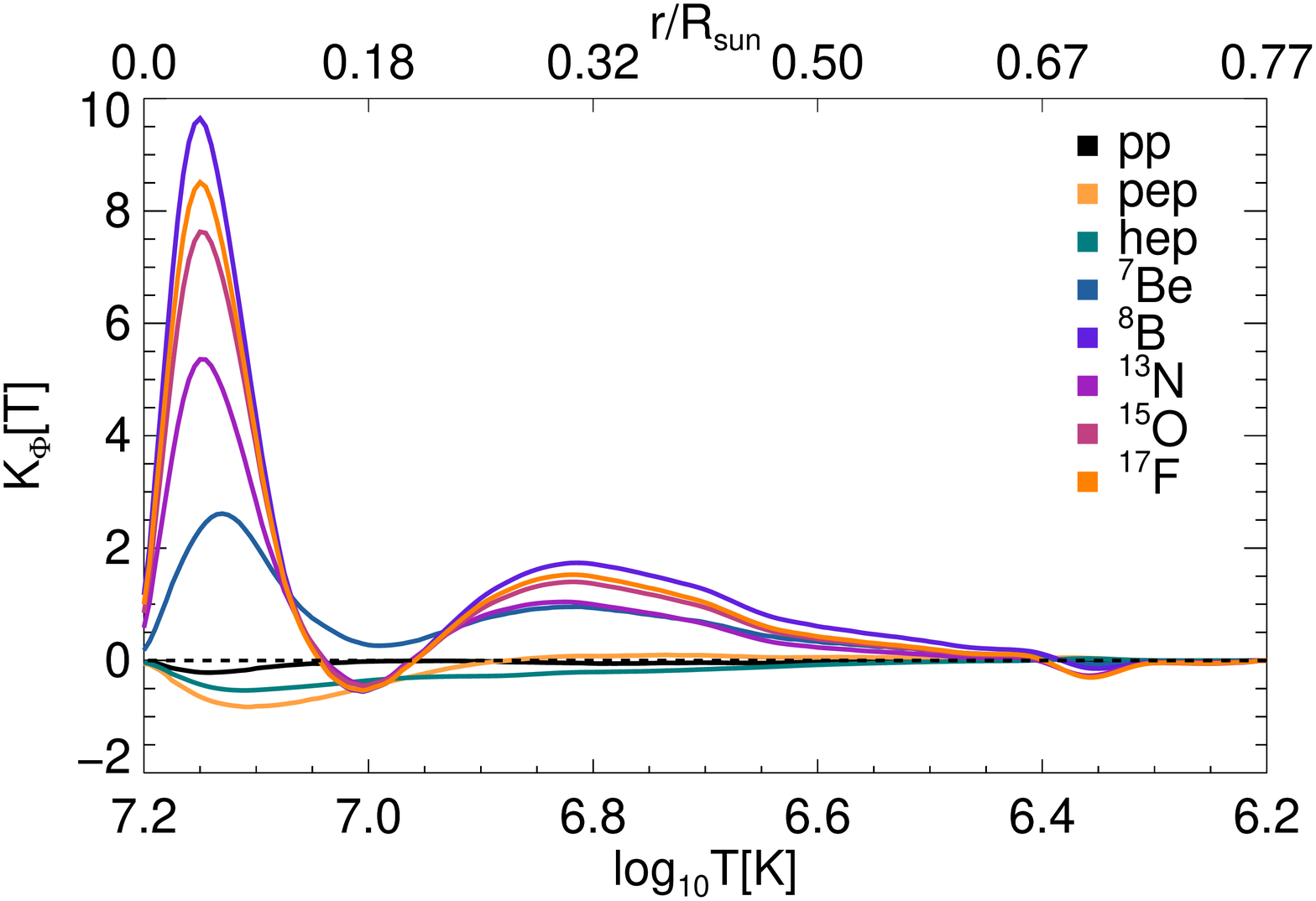}	
	\includegraphics[width=0.49\columnwidth]{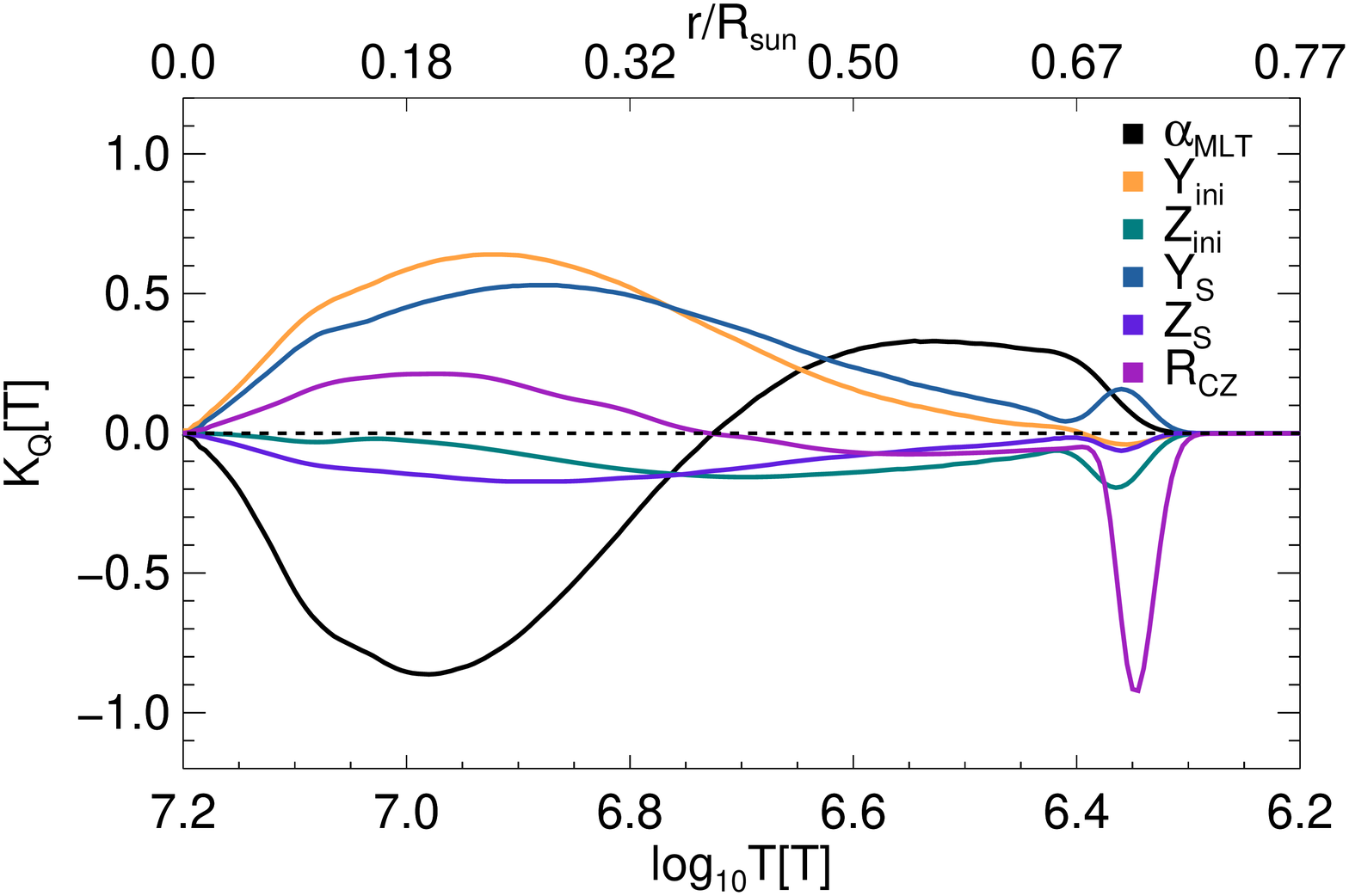}	
        \caption{Opacity kernels for different quantities as a function of the solar temperature. Top-left panel: Five points of the sound speed at different solar radius; $c1=c(0.06 R_\odot)$, $c2=c(0.15 R_\odot)$, $c3=c(0.28 R_\odot)$, $c4=c(0.48 R_\odot)$ and $c5=c(0.66 R_\odot)$. Top-right panel: Neutrino fluxes. Low panel: Initial and surface metallicity and helium, $\alpha_{MLT}$ and radius at the base of the convective envelope.}
        \label{fig:kernels}
\end{figure}

\begin{figure}
	\centering 
    \includegraphics[width=0.49\columnwidth]{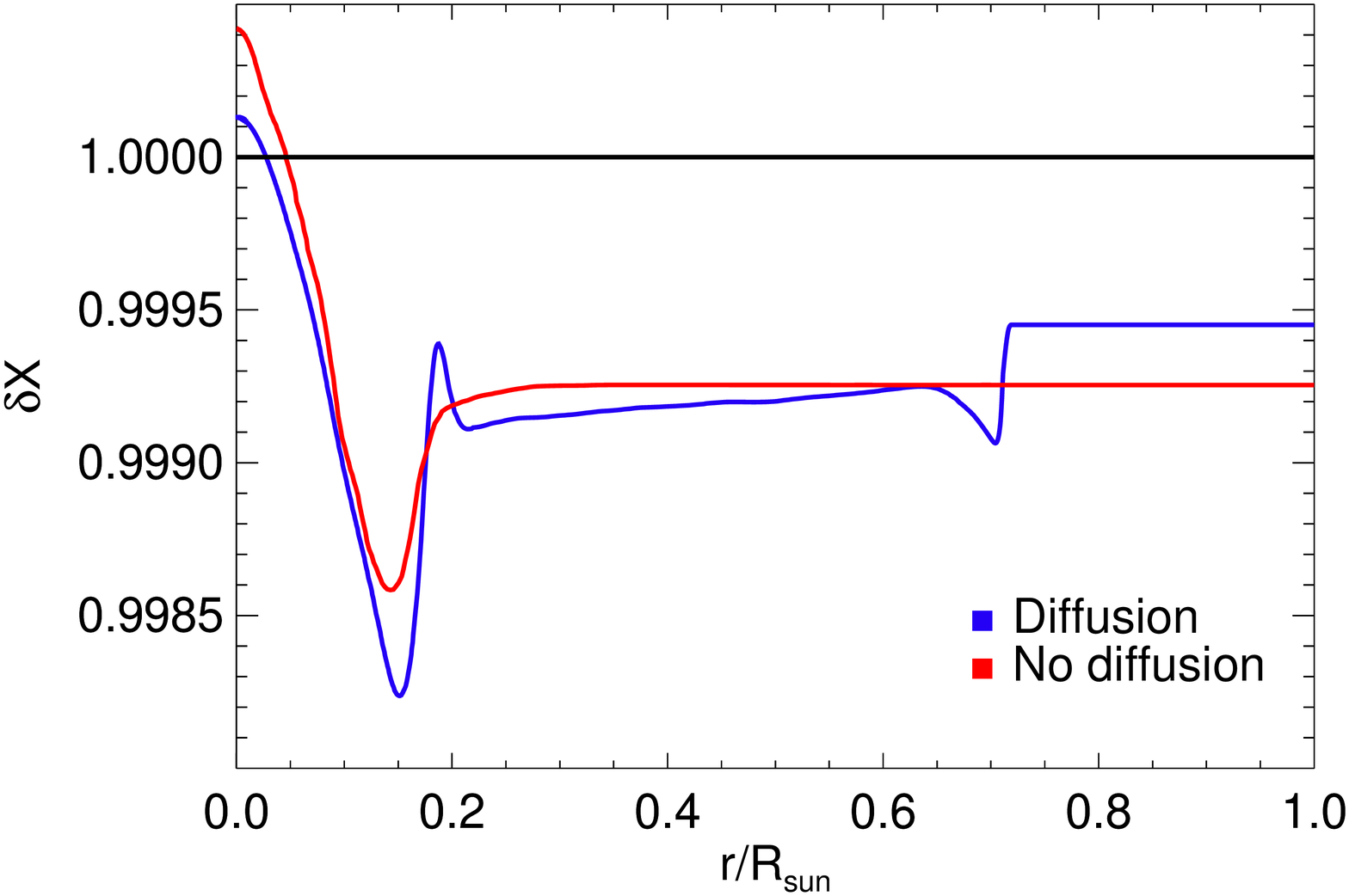}
        \caption{Relative changes for the hydrogen abundance profile when an opacity perturbation at $\log{T_0} = 7.0$ (corresponding to $r=0.18R_\odot$) is introduced. The solid red line corresponds to a SSM model with diffusion and the solid blue line to a SSM without diffusion.}
        \label{fig:diffusion}
\end{figure}

In Fig.\,\ref{fig:kernels}, the kernels are plotted as function of the temperature for the different relevant outputs of the system.
 Comparing our results with the ones presented in \cite{Tripathy98}
 and \cite{villante10}, we find a very good agreement with the present
 calculations\footnote{Note that \cite{villante10} adopts an
alternative approach in which the solar structure equations are first
linearized and then solved and gives the opacity kernels as function
of the solar radius.}.
The only noticeable difference with \cite{Tripathy98} 
is found for the kernel of the hydrogen abundance. Note, however, 
that our SSMs are calculated using diffusion while it was not taken into account in \cite{Tripathy98}. 
In Fig.\,\ref{fig:diffusion}, we present the relative changes for the hydrogen abundance profile when the opacity is perturbed at $\log_{10}{T_0} = 7$ with and without diffusion included.

\bibliography{ssm}

\end{document}